\documentclass[11pt]{article}
\usepackage[margin=1in]{geometry}
\usepackage{setspace}
\usepackage{enumitem}
\usepackage{float}
\usepackage{amsmath}
\usepackage{amsfonts}
\usepackage[none]{hyphenat}
\usepackage{amssymb}
\usepackage{latexsym}
\usepackage{graphicx}
\usepackage[english]{babel}
\usepackage{cite}
\usepackage{tocloft}
\usepackage{tikz}
\usepackage{hyperref}
\hypersetup{
    colorlinks=true,
    linkcolor=blue, % Change this color to your desired color
    citecolor=orange, % Change this color to your desired color
    urlcolor=teal   % Change this color to your desired color
}
\addto\captionsenglish{
  \renewcommand{\contentsname}%
    {Contents}%
    \vspace{-1.3em}
}
\begin{document}
\input epsf

\def\p{\partial}
\def\h{{1\over 2}}
\def\be{\begin{equation}}
\def\bea{\begin{eqnarray}}
\def\ee{\end{equation}}
\def\eea{\end{eqnarray}}
\def\d{\partial}
\def\la{\lambda}
\def\eps{\epsilon}
\def\bb{\bigskip}
\def\mm{\medskip}
\def\beq{\begin{equation}}
\def\eeq{\end{equation}}
\newcommand{\dm}{\begin{displaymath}}
\newcommand{\edm}{\end{displaymath}}
\renewcommand{\b}{\tilde{B}}
\newcommand{\gm}{\Gamma}
\newcommand{\ac}[2]{\ensuremath{\{ #1, #2 \}}}
\renewcommand{\ell}{l}
\newcommand{\z}{\ell}
\newcommand{\newsection}[1]{\section{#1} \setcounter{equation}{0}}
\def\bb{$\bullet$}
\def\Qbar{{\bar Q}_1}
\def\QPbar{{\bar Q}_p}

\def\q{\quad}

\def\bn{B_\circ}

\def\MM#1{{\bf \textcolor{red}{MM:} }{\textcolor{blue}{#1}}}
\def\h{{1\over 2}}
\def\t{\tilde}
\def\r{\rightarrow}
\def\nn{\nonumber\\}
\let\bm=\bibitem
\def\Kt{{\tilde K}}
\def\b{\bigskip}

\let\p=\partial

\newcommand\blfootnote[1]{%
  \begingroup
  \renewcommand\thefootnote{}\footnote{#1}%
  \addtocounter{footnote}{-1}%
  \endgroup
}
\numberwithin{equation}{section}
\newcounter{daggerfootnote}
\newcommand*{\daggerfootnote}[1]{%
    \setcounter{daggerfootnote}{\value{footnote}}%
    \renewcommand*{\thefootnote}{\fnsymbol{footnote}}%
    \footnote[2]{#1}%
    \setcounter{footnote}{\value{daggerfootnote}}%
    \renewcommand*{\thefootnote}{\arabic{footnote}}%
    }

\begin{flushright}
\end{flushright}
\vspace{20mm}
\begin{center}
{\LARGE The universal thermodynamic properties of Extremely Compact Objects
 }

\vspace{18mm}
{\bf Samir D. Mathur$^{1}$ and Madhur Mehta$^2$
}

\blfootnote{$^{1}$ email: mathur.16@osu.edu }
\blfootnote{$^{2}$ email: mehta.493@osu.edu}

\vspace{4mm}

\b

Department of Physics

 The Ohio State University
 
Columbus,
OH 43210, USA

\b

\vspace{4mm}
\end{center}
\vspace{10mm}
\thispagestyle{empty}
\begin{abstract}

An extremely compact object (ECO) is defined as a quantum object without horizon, whose radius is just a small distance $s$ outside  its Schwarzschild radius. We show that any ECO of mass $M$ in $d+1$ dimensions with $s\ll (M/m_p)^{2/(d-2)(d+1)}l_p$ must have (at leading order) the same thermodynamic properties --- temperature, entropy and  radiation rates  --- as the corresponding semiclassical black hole of mass $M$. An essential aspect of the argument involves showing that the Tolman-Oppenheimer-Volkoff equation has no consistent solution in the region just outside the ECO surface, unless this region is filled with radiation at the (appropriately blueshifted) Hawking temperature.  In string theory it has been found that black hole microstates are fuzzballs --- objects with no horizon --- which are expected to have  a radius that is only a little larger than  the horizon radius. Thus the arguments of this paper provide a nice closure to the fuzzball paradigm:  the absence of a horizon removes the information paradox, and the thermodynamic properties of the semiclassical hole are nonetheless recovered to an excellent approximation.

\end{abstract}
\vskip 1.0 true in

\newpage

\tableofcontents
\pagenumbering{gobble}
\section{Introduction}

Consider a ball of steel with mass $M$. This mass does not determine the temperature $T$ of the ball; we can choose different values of the temperature, including $T=0$. There is also no simple relation between the mass $M$ and the entropy $S$ of the ball; this entropy depends on the detailed composition of the ball. Even if we know $M, T, S$, we cannot predict the radiation rate $\Gamma$ from the ball; this radiation rate depends on the surface area and shape of the ball. 

Remarkably, the situation is much simpler for black holes. Consider the Schwarzschild hole in $3+1$ dimensional spacetime. Hawking \cite{hawking} found that the temperature of the hole is
\be
T_\text{H}={1\over 8\pi GM}\, .
\label{one}
\ee
The entropy is \cite{hawking,bek}
\be
S_{bek}={A\over 4G}\, ,
\label{two}
\ee
where $A$ is the surface area of the horizon. Hawking found that the hole radiates `thermally' in the sense that the radiation rate is related to  the absorption cross section in the manner expected from  detailed balance. Thus consider  the radiation of a massless scalar field. Then the number of particles emitted per unit time in the energy range $d\omega$ is $\Gamma_\text{H}[l,m,\omega]{d\omega\over 2\pi}$ with
\be
\Gamma_\text{H}[l,m,\omega]= {{\cal P}[l,m,\omega]\over e^{\omega\over T_\text{H}}-1} \, ,
\label{three}
\ee
where ${\cal P}[l,m,\omega]$ is the absorption probability for an incoming spherical wave of energy $\omega$ in the spherical harmonic $Y_{l,m}$. 

Thus, the thermodynamical properties of the Schwarzschild hole are determined completely by the mass $M$ of the hole. More generally, the thermodynamical properties  of the hole are determined by the conserved quantum numbers characterizing the hole---the mass $M$, the charges $Q_i$ and the angular momenta $J_i$. What is the reason for this very special behavior of black holes?

 The universal thermodynamical behavior of black holes is often attributed to the presence of a {\it horizon} in the black hole geometry. Bekenstein's argument for entropy \cite{bek} started with the idea that the entropy of matter falling through a horizon is `lost' to the outside world; to prevent a violation of the second law of thermodynamics one must then attribute an entropy to the hole, which ends up taking the value (\ref{two}). One would not make such an argument for the entropy of matter falling into a normal box, since we would not think of this entropy as having been `lost'; thus the existence of  a horizon was important to the argument. Similarly, Hawking's computation of radiation from the hole, which leads to the relations (\ref{one}) and (\ref{three}), involves the horizon in fundamental way. Outgoing null geodesics in the vicinity of the horizon separate: the ones just outside the horizon ultimately escape to infinity, while those just inside the horizon fall into the singularity. This separation leads to a stretching of spacelike slices in the black hole geometry, and the consequent production of particle pairs around the horizon. One member of the pair (which we call $b$) escapes to infinity as `Hawking radiation', while the other member (which we call $c$) falls into the hole with negative energy and thus lowers the mass of the hole. 
 \pagenumbering{arabic}
 The above observations indeed suggest that there is a close relation between the existence of a horizon and the emergence of black hole thermodynamics (\ref{one})-(\ref{three}). There have been  attempts to relate the fact that a horizon `traps information inside' to the notion that entropy is a `lack of information'.
 
  But this traditional picture of the hole possessing a horizon leads to a serious problem---the information paradox \cite{hawking, Hawking:1976ra}. The $(b,c)$ quanta of the Hawking pair are in an entangled state, which can be schematically written as
 \be
 |\psi\rangle_{pair}={1\over \sqrt{2}} \left ( |0\rangle_b|0\rangle_c+|1\rangle_b|1\rangle_c\right )\,.
 \ee
Thus we get a monotonically rising entanglement between the emitted radiation and the remaining hole, leading to a sharp puzzle near the endpoint of evaporation. If the hole evaporates away completely, then the radiation is left in an entangled state, but there is nothing that it is entangled {\it with}; such radiation cannot be described by any quantum state, leading to a violation of quantum unitarity. If the evaporation terminates in a planck sized remnant, then we face difficulties with a planck size object having an unbounded number of internal states. The small corrections theorem \cite{cern} shows that Hawking's argument is stable against any small correction to the horizon dynamics; we cannot escape the problem of monotonically rising entanglement by seeking subtle correlations among the large number of radiated quanta.

String theory computations suggest that the microstates of the black hole are horizon sized quantum objects called {\it fuzzballs} \cite{Lunin:2001jy, Lunin:2002iz, Kanitscheider:2007wq, Mathur:2005zp, Bena:2007kg, Chowdhury:2010ct, Bena:2015bea, Bena:2016ypk, Mathur:2020ely, Bena:2022rna}. A fuzzball does not have a horizon and radiates from its surface like any other normal body; thus there is no information paradox. An entropic argument indicates that the surface of a generic fuzzball should be at a proper distance $s\sim l_p$ outside the Schwarzschild radius $r_0$ of the semiclassical hole \cite{ghm}. Thus fuzzballs are  expected to be extremely compact. In what follows we will use the term ECO to represent any `Extremely Compact Object'; i.e., an object with no horizon and a radius which is just a little larger than the horizon radius $r_0$ for its mass $M$.
 
 In this paper we will argue that any Extremely Compact Object (ECO) will have the {\it same} thermodynamical properties (\ref{one})-(\ref{three}), as the semiclassical black hole, to leading order.  For $d+1$ dimensions, our arguments will hold for ECOs whose surface is at a proper distance
  \be
 s_\text{ECO}\ll \left ( {M\over m_p}\right )^{2\over (d-2)(d+1)} l_p\,,
 \label{seven}
 \ee
 outside the horizon radius $r_0$.
 For $3+1$ dimensions, this is
 \be
 s_\text{ECO}\ll \left ( {M\over m_p}\right )^\h l_p\,.
 \label{threeq}
 \ee
The deviations of the thermodynamics parameters of the ECO from the corresponding values of the black hole become smaller as $s$ is made smaller. The essence of the argument will rely on looking at the quantum stress tensor just outside the surface of the ECO   (a brief version of this argument was presented in the essay \cite{Mathur:2023uoe}).

Before proceeding with our analysis, we note a result obtained in \cite{israel,israel2,Lemos:2009uk,Lemos:2010kw,burman2024smooth}. Suppose the black hole is replaced by a thin spherical shell which is supported by its own pressure, and stands a small distance outside its horizon radius. It was argued that such a shell will have to be in equilibrium with the  local Unruh radiation, and that this fact leads to the entropy (\ref{two}) for the shell. This shell thus gives a `brick wall' type of model \cite{thooft, banerjee2024brickwall}  for a black hole, which is at the standard black hole temperature $T_\text{H}$. In the present paper, our interest is in the converse question: can there be an ECO whose temperature $T_\text{ECO}$ is {\it different} from $T_\text{H}$? We will consider a general ECO, rather than any particular model like a thin shell.  Thus our arguments will proceed on somewhat different lines from the arguments of \cite{israel}, and will not involve the internal structure of the ECO at all.    But like \cite{israel} we will also use  the fact that the near-surface region of an ECO has a negative vacuum energy, equal in magnitude to  the energy of Unruh radiation near a black hole horizon.

 We also note that many aspects of compact objects   have been  studied in \cite{cardoso,mann,Lemos:2008vu, Lemos:2017mci,Fernandes:2022gjd,Brustein:2023hic, afshordi, wang}.

\subsection{Plan of the paper}

The plan of this paper is as follows:

\begin{enumerate}[label =(\roman*)]

\item  In section\,\ref{sec2} we describe the structure of an ECO, and state conditions ECO1-ECO3 which will define an ECO.

\item In section\,\ref{sec3} we note that the vacuum state near the surface of an ECO has a negative energy density, which (to leading order) is the same as the negative energy density of the Boulware vacuum.

\item In section\,\ref{sec4} we note that if the ECO had the same temperature $T_\text{ECO}(M)$ as the temperature $T_\text{H}(M)$ for a black hole then we will also have an agreement of entropies $S_\text{ECO}(M)=S_{bek}(M)$; this just follows from the laws of statistical mechanics valid for systems with many degrees of freedom.
We then  show that if $T_\text{ECO}=T_\text{H}$, then the radiation rate from the ECO, $\Gamma_\text{ECO}[\{ l\}, \omega]$, will also be equal (to leading order) to  the radiation rate from  the black hole, $\Gamma_\text{H}[\{ l\}, \omega]$. This argument relies on the extreme compactness of the ECO.

\item In section \ref{sec5} we give a heuristic derivation of the result that the temperature of an ECO must equal the temperature of the black hole.  The arguments will be rough but they nonetheless capture the essential physics of the problem and lead to the scale (\ref{seven}) for compactness of an ECO.

\item In section \ref{sec6} we will make a plausible assumption about the stress tensor describing the vacuum near the ECO surface, and solve the Tolman-Oppenheimer-Volkoff (TOV) equation in a near-surface approximation. We will find that if the ECO satisfies our conditions ECO1-ECO3, then there is no consistent solution to this equation if $T_\text{ECO}\ne T_\text{H}$.

\item In section \ref{sec7} we consider   $1+1$ dimensional dilaton gravity coupled to minimal scalars. This case was analyzed in detail in \cite{PhysRevD.105.025015}, and we will recall their results. In this analysis the stress tensor of the vacuum is taken into account exactly, since it can be computed from the conformal anomaly. We note similarities and differences between this $1+1$ dimensional case and the behavior in higher dimensions. 

\item Section \ref{sec8} is a summary and discussion.

\end{enumerate}

\section{Structure of an ECO}\label{sec2}

In this section we will  describe the structure of what we mean by an Extremely Compact Object (ECO). While describing this structure we will extract properties ECO1-ECO3 which will define  our  ECO. 

We will assume for simplicity that the ECO is spherically symmetric to leading order. Thus while the interior of the ECO can have structures at microscopic scales that are not  spherically symmetric, we will assume that the  gravitational properties of the ECO in the region outside the ECO are well approximated by a spherically symmetric and time-independent configuration. The ECO is described by a mass $M$ measured at infinity, and has vanishing angular momentum  and charge. 

The number of space dimensions will be called $d$. We define the planck length as
\be
G=l_p^{d-1}\,,
\ee
and the planck mass as
\be
m_p={1\over l_p}\,.
\ee

We will assume that the mass $M$ of the ECO satisfies
\be
{M\over m_p}\gg 1\,.
\ee

\subsection{The radius $R_{\text{ECO}}$, and the condition ECO1}

Our Extremely Compact Object should be characterized by a radius $R_\text{ECO}$ that is just a little larger than the horizon radius for a black hole of mass $M$. In such a situation, the surface of the ECO feels a strong  inward pull of gravity, and for simple distributions of matter, an equilibrium solution is not possible. For example the Buchdahl theorem\cite{buchdahl}  in 3+1 dimensions says that a spherically symmetric ball of perfect fluid (with pressure decreasing outwards) cannot resist collapse if its radius  is less than ${9\over 4}GM$. In string theory, it was noted in \cite{Mathur:2016ffb} how fuzzballs evade such a conclusion because the compact dimensions are not trivially tensored with the noncompact ones. Recently, string theory constructions have been given for solitonic stars; formally these structures can be extrapolated to a point where they are only slightly larger than their Schwarzschild radius \cite{Bah:2022yji, Heidmann:2021cms, Heidmann:2023thn, Heidmann:2023kry, Bah:2023ows}. 

Our ECO will be described by a radius $r=R_\text{ECO}$. Here the coordinate $r$ is is defined in the usual way as the radius where the area of the angular sphere is $4\pi r^2$ in 3+1 dimensions, and $r^2\Omega_{d-1}$ in $d+1$ dimensions. We will make no assumptions about the structure in the region $r<R_\text{ECO}$; this can be a region with significant quantum gravitational effects, so that spacetime itself may not make sense here. But in the region $r>R_\text{ECO}$ we require usual semiclassical dynamics to hold to sufficient accuracy for all our purposes. 
In the semiclassical region $r>R_\text{ECO}$ we take the metric ansatz in $d+1$ spacetime dimensions
\be
ds^2=-e^{2\alpha(r)} dt^2 + e^{2\beta(r)} dr^2 + r^2 d\Omega_{d-1}^2\,.
\label{five}
\ee
 The Einstein equations $G_{\mu\nu}=8\pi GT_{\mu\nu}$ have the vacuum Schwarzschild solution in 3+1 dimensions
\be
ds^2=-\left(1-{2GM\over r}\right)dt^2+{dr^2\over 1-{2GM\over r}}+r^2 d\Omega_2^2\,.
\label{fiveq}
\ee
For d+1 dimensions the vacuum Schwarzschild solution is
\be
ds^2=-\left(1-{r_0^{d-2}\over r^{d-2}}\right)dt^2 + {dr^2\over 1-{r_0^{d-2}\over r^{d-2}}}+r^2d\Omega_{d-1}^2\, ,
\label{six}
\ee
where
\be
r_0^{d-2}=\mu GM, ~~~\mu={16\pi\over (d-1)\Omega_{d-1}}\,.
\label{mseven}
\ee
Here and in all later equations, when we talk of a general space-time dimension $d+1$, we have in mind $d\ge 3$. In $2+1$ dimensional gravity the mass $M$ does not generate an asymptotically flat spacetime, and we will not consider this case here. The case of $1+1$ dimensional dilaton gravity  will be considered separately  in section\,\ref{sec7}; this case allows for a more explicit solution than is possible in higher dimensions.

The requirement that the ECO be extremely compact says that $R_\text{ECO}$ should be only a little larger than the horizon radius for the same mass $M$. For the black hole metric (\ref{six}), we can describe the near-horizon region using Rindler coordinates. 
 We define
\be
\t t =  {(d-2) \over 2r_0}\, t, ~~~~s =2\left ( {r_0(r-r_0)\over d-2}\right )^\h\,.
\label{knine}
\ee
 The near-horizon geometry then becomes Rindler space
\be
ds^2\approx -s^2 d{\t t}^{\,2} + ds^2 + dx_1^2+\dots + dx_{d-1}^2\,,
\label{sixqq}
\ee
with $x_1\dots  x_{d-1}$ describing the tangent space to the angular sphere. 

It is convenient to recast the compactness requirement  on $R_\text{ECO}$ as follows. The horizon radius for a black hole of mass $M$  is $r=r_0$. Let $s_\text{ECO}$ be the proper distance, measured radially, between a sphere with radius $r=r_0$ and a sphere with radius $r=R_\text{ECO}$, in the black hole metric (\ref{six}). From (\ref{knine}) we see that
\be
s_\text{ECO}=2\left ( {r_0(R_\text{ECO}-r_0)\over d-2}\right )^\h\,.
\label{eq.sECO}
\ee
An ECO should be characterized by 
$s_\text{ECO}\ll r_0$. 
  For fuzzballs, entropic arguments indicate that $s_\text{ECO}\sim l_p$. The arguments of the present paper will hold for a larger range of $s_\text{ECO}$ (\ref{seven}); this range will emerge in the course of our analysis. Note that in defining $s_\text{ECO}$ we use the black hole metric (\ref{six})  only as a convenient tool to describe the difference between the radius values $r=R_\text{ECO}$ and $r=r_0$; the actual metric in the region around $r=R_\text{ECO}$ can be very different from the vacuum black hole solution.

We summarize the above discussion in the following property of the ECO:

\b

{\bf ECO1:} \quad Semiclassical physics holds outside the radius $r=R_\text{ECO}$, and this radius  $R_{\text{ECO}}$ is close to the horizon radius $r_0$, with 
 \be
 s_\text{ECO}\ll \left ( {M\over m_p}\right )^{2\over (d-2)(d+1)} l_p.
 \label{sevenq}
 \ee

\b

We can think of (\ref{sevenq}) as describing the  `compactness' of  an ECO. In what follows we will sometimes refer to the scale 
\be
s_c\, \sim\,  \left ( {M\over m_p}\right )^{2\over (d-2)(d+1)} l_p\, \sim\,  \left ( {r_0\over l_p}\right)^{2\over d+1} l_p\,,
\label{zzel}
\ee
 as the compactness scale.

\subsection{Redshift at the ECO surface, and the condition ECO2}\label{sectemp}

The essential property of a black hole is the infinite redshift that we get at the black hole horizon. Any extremely compact object that replaces a black hole should be characterized by a very large (though not infinite) redshift at its surface $R_{\text{ECO}}$. Let us note what the scale of this redshift should be.

We define the redshift parameter $q(r)$ by
\be
q(r)\equiv (-g_{tt}(r))^{-\h}\,.
\label{zztw}
\ee
In the Rindler region $(r-r_0)/r_0\ll 1$, for the black hole metric (\ref{six}), we have
\be
q(r)\approx \left({(d-2)(r-r_0)\over r_0} \right)^{-\h}\approx {2r_0\over (d-2) s}\,.
\label{kone}
\ee
At the compactness scale $s_c$ (eq.(\ref{zzel})), this redshift parameter is
\be
q(s_c)\sim \left( {M\over m_p}\right )^{-{2\over (d-2)(d+1)}} \frac{r_0}{l_p}\sim \left( { r_0\over l_p}\right)^{d-1\over d+1}\sim \left ( { M\over m_p}\right)^{(d-1)\over (d-2)(d+1)}\, .
\ee
Since the ECO is required to have $s_{\text{ECO}}\ll s_c$, we place the following requirement on our ECO:

 \b
 
 {\bf ECO2:} \quad The redshift at $r=R_\text{ECO}$ satisfies
\begin{align}
    q(R_{\text{ECO}})\gg  \left(\frac{r_0}{l_p}\right)^{\frac{d-1}{d+1}}\sim \left ( { M\over m_p}\right)^{(d-1)\over (d-2)(d+1)}
\label{ksixq}\,.
\end{align}

\subsection{Small energy outside the ECO, and the condition ECO3}\label{secenergy}

It may appear reasonable to require that the spacetime in the region $r>R_\text{ECO}$ has zero stress-energy and thus a metric of the black hole form (\ref{six}). But the ECO has in general a nonzero temperature $T_\text{ECO}$, 
and radiation corresponding to this temperature will fill up the region near the ECO. We have in mind temperatures of order the Hawking temperature (\ref{one}), and it is true that the radiation energy density at this temperature is very small at $r\gg R_\text{ECO}$. But the energy density of this radiation can be very large close to $R_\text{ECO}$ due to the large redshift in this region, and contribute a total mass that is $O(M)$. Thus we have to consider the more general ansatz (\ref{five}) for the metric in the region just outside $r=R_\text{ECO}$.

To clarify this point, let us estimate the distance $s$ from the horizon radius $r_0$ upto which this radiation density is appreciable. Suppose the temperature of the ECO as seen at infinity is $T_\text{ECO}$. The radiated quanta suffer a redshift as they move out of the gravitational potential of the ECO. Thus the effective temperature of the radiation at a radius $r$, measured in a local orthonormal frame with time direction along $t$, is 
\be
T_{\text{ECO}}(r)=q(r)T_\text{ECO}\,,
\ee
where $q(r)$ is the redshift parameter (\ref{zztw}). 
Using (\ref{kone}) for the value of $q(r)$ in the Rindler region $(r-r_0)/r_0\ll 1$, for the black hole metric (\ref{six}), we find that the temperature at a distance $s$ from the horizon radius $r_0$ is
\be
T_{\text{ECO}}(s)\approx {2r_0\over (d-2) s}T_\text{ECO}\,.
\ee
Assuming that $T_\text{ECO}$ is of the same order as the Hawking temperature $T_\text{H}\sim r_0^{-1}$, we have
\be
T_{\text{ECO}}(s)\sim {1\over s}\,.
\ee
Thus we see that at small $s$, the local temperature is very high, and in fact this temperature reaches planck scale at planck distance $s\sim l_p$ from the horizon radius. The energy density of a massless quantum field at temperature $T$ is
\be
\rho=a T^{d+1}\,,
\label{mone}
\ee
where $a$ is a constant of order unity, depending on the dimension $d$ and the spin of the quantum field. For a scalar field in 3+1 dimensions
\be
a={\pi^2\over 30}\,.
\ee

If there is nonzero stress energy outside $R_\text{ECO}$, then how should we capture the compactness of the ECO through the condition (\ref{sevenq}) on $R_\text{ECO}$? In particular, there is a singular situation that formally satisfies the conditions ECO1, ECO2, but which we should exclude from our consideration of ECOs. As we will discuss later in section\,\ref{sec6}, the Einstein equations allow a `truncated isothermal star', which has the following structure. There is an isothermal photon gas filling the region $0<r<R$, with the matter density truncated in some way at $r=R$. The density rises to infinity at $r\r0$, and the redshift $q(r)$ also diverges at $r\r 0$. Thus one could place a surface just outside $r=0$ and find a high redshift at this surface, satisfying condition ECO2. The surface also has a very small radius, so it would formally satisfy the spirit of the requirement ECO1. But such an object is not what we would consider an ECO for our purposes, since almost all its mass is outside the high redshift surface that we have defined. Thus we would like to impose a condition on our ECO which says that there is very little matter outside the compactness scale $s_c$

Note that the energy density of thermal radiation will typically fall off as a power law as we go out from the ECO. Thus we cannot ask that it be exactly zero at some given distance from the ECO surface, but we can ask that its effects not be relevant if we are sufficiently far from the ECO surface.   We do this by requiring that outside the compactness scale $s_c$, the geometry is close to the metric of the black hole:

\b

{\bf ECO3:} \quad At distances $s>\left ( {M\over m_p}\right )^{2\over (d-2)(d+1)} l_p$ from the black hole horizon radius $r_0$, the geometry is well approximated by the black hole metric (\ref{six}).

\b

In particular, at distances $s>s_c$ from the black hole radius $r_0$, the redshift factor is of order the redshift for the black hole metric
\be
q_{\text{ECO}}(r)\sim \left (1-{r_0^{d-2}\over r^{d-2}}\right )^{-\h}\, .
\label{zzten}
\ee

\b

In section\,\ref{seccheck} we will check that it is consistent to impose condition ECO3; i.e., we will verify that the stress-tensor of the thermal radiation near the ECO surface does not significantly distort the black hole metric at distances $s\gtrsim s_c$ from the horizon radius $r_0$.

\subsection{A relation following from condition ECO1}

The condition ECO1 says that semiclassical dynamics is a good approximation at $r>R_\text{ECO}$. From the discussion of section\,\ref{secenergy}, we see that there will in general be a nonvanishing energy density  $T^{t}{}_{ t}=-\rho$ in this region.  We can then use the ansatz (\ref{six}) for the metric, and solve the equation $G_{tt}=8\pi G T_{tt}$ with this energy density. In 3+1 dimensions we get
\be
e^{-2\beta(r)}=1-{2GM(r)\over r}\,.
\label{ccfone}
\ee
For a star, we have
\be
M(r) = \int_{0}^r dr\, 4\pi r^2 \rho(r)\,.
\ee
For an ECO,  the region $0<r<R_\text{ECO}$  can have large quantum gravitational effects, and thus may not be well approximated as a smooth manifold. Thus we do not wish to integrate over $r$ in this region. But we can compute $M(r)$ by integrating the mass density {\it outside} the ECO, using the fact that the mass as seen from infinity is $M$:
\be
M(r)=M-\int_r^\infty dr\,4\pi r^2 \rho(r)\,.
\ee
In $d+1$ dimensions, we have
\be
e^{-2\beta(r)}=1-{\mu G M(r)\over r^{d-2}}, ~~~\mu={16\pi\over (d-1)\,\Omega_{d-1}}\,,
\label{ktwo}
\ee
with
\be
M(r)=M-\int_r^\infty dr\, \Omega_{d-1} r^{d-1} \rho(r)\,.
\label{ktwop}
\ee
The regularity of the ECO solution  thus yields the following requirement. In the region $r>R_\text{ECO}$ we must have everywhere
\be\label{eq.ECO4}
1-{\mu G M(r)\over r^{d-2}}>0\,.
\ee

\section{The vacuum stress-energy near an ECO}\label{sec3}

In section \ref{secenergy} we talked about the energy density of radiation outside the ECO. But even if our object has a temperature $T=0$, there is still an important energy to consider outside the surface of the object. There is a nonvanishing  vacuum stress-energy of quantum fields in the region around the object, caused by the behavior of field modes in the metric created by the object. As we will note below, for temperatures which are of order the Hawking temperature, this vacuum stress-energy is of the same order as the stress-energy of the radiation near the surface. Thus this vacuum energy will play an important role in our analysis.

We will begin by recalling the computation of vacuum energy for the Schwarzschild hole. We will then note that an ECO has, to leading order, the {\it same}  vacuum energy as a black hole. Taking into account the thermal radiation just outside the ECO, we will obtain the total stress energy in this near-surface region.

\subsection{The vacuum stress-tensor near a black hole horizon}

The vacuum energy in a spacetime depends in general on the choice of the quantum state for the fields. In a black hole spacetime, some  commonly considered states are the Unruh vacuum, the Hartle-Hawking vacuum and the Boulware vacuum. In 3+1 dimensions the vacuum energy for these states was computed in \cite{candelas}, using methods developed in \cite{Unruh:1976db, Boulware:1974dm, Hartle:1976tp, Christensen:1976vb, Fulling:1977zs, Christensen:1977jc}. This computation is in general quite complicated, but there is a simple way to get the answer for the quantity we need: the vacuum stress energy for the Boulware vacuum to leading order close to the horizon.

Let us first recall how the computation of the vacuum stress-energy proceeds in  general for the black hole metric (\ref{six}). Consider for simplicity a scalar field $\hat\phi$ satisfying $\square \hat\phi=0$. We expand $\hat \phi$ 
 in modes in the region $r>r_0$
\be
\hat\phi= \sum_{\{ l\},k} \Big (~   \hat a_{\{ l\},k}\, f_{\{ l \},k}(r) Y_{\{ l\}}(\{ \Phi\}) e^{-i\omega_{\{ l \},k}\, t}+\hat a^\dagger_{\{ l \},k}\,  f^*_{\{ l \},k}(r) Y^*_{\{ l \}}(\{ \Phi \} ) e^{i\omega_{\{ l \},k}\, t}  ~\Big )\,,
\label{el}
\ee
where $\{ l \}$ denote the quantum numbers of the angular harmonic and $\{ \Phi \}$ denote the angular variables on $\Omega_{d-1}$. 
 The effective potential  felt by these modes $f_{\{ l \},k}(r)$  is sketched in fig.\ref{ECOscalarpotential}(a). The potential vanishes near the horizon, and also at infinity, with a barrier in the region in between. 
 
 Using the above modes, one computes the correlator 
 \be
 \langle \Psi| \hat \phi(x)\hat\phi(y)|\Psi\rangle\,,
 \ee
 for infinitesimally separated $x, y$. The computation of this correlator requires knowledge of the quantum state $|\Psi\rangle$ of the field $\hat\phi$. We then find the expectation value of the stress-energy tensor by computing taking  appropriate derivatives in $x,y$,  subtracting a normal ordering constant, and then taking the limit $x\r y$. 
 
 Knowledge of the above procedure will be useful to us below, but for now we note that for the leading order stress-tensor in the Boulware vacuum, there is an easier way to obtain the result. The argument proceeds as follows. In the Hartle-Hawking vacuum, the black hole is in equilibrium with its radiation. The geometry is smooth at the horizon,  and there is no flux into or out of the horizon. Thus to leading order the stress-energy tensor is zero around the horizon. (Here by `leading order' we mean that  we are ignoring contributions of order $\langle T^\mu{}_\nu\rangle \sim r_0^{-(d+1)}$ which arise from the energy density of quanta with wavelength $\lambda\sim r_0$, and terms arising from the anomaly; these terms are regular at the horizon.) We can understand this vanishing of the stress energy by going to Rindler coordinates (\ref{knine}) near the horizon. Then the Hartle-Hawking vacuum is like the Minkowski vacuum, which has a vanishing stress tensor.  
 
 Now consider the Boulware vacuum $|0\rangle_\text{B}$. This vacuum state is obtained by requiring
 \be
  \hat a_{\{ l\},k}|0\rangle_\text{B}=0\,,
  \ee
  for the operator modes in (\ref{el}). To understand the nature of this state, consider the near horizon Rindler region. In this region   the modes $f_{\{ l \},k}(r)$ in (\ref{el}) are Rindler modes and the Boulware vacuum is the Rindler vacuum. In Rindler coordinates, the Minkowski vacuum appears to be populated with Rindler excitations at a temperature
  \be
  T(s)={1\over 2\pi s}\,.
  \label{mtwo}
  \ee
  Since the Minkowski vacuum has vanishing stress tensor, the stress tensor of the Rindler vacuum is given by the {\it negative} of the stress tensor of thermal radiation at temperature (\ref{mtwo}). 
  
  The Hawking temperature of the black hole in $d+1$ dimensions is
  \be
T_\text{H}={(d-2)\over 4 \pi  r_0}\,,
\label{msix}
\ee
and (\ref{mtwo}) is just this temperature times the redshift factor at distance $s$ from the horizon. In terms of the coordinate $r$, 
   the local temperature  $T_\text{H}(r)$ is 
  \be
T_\text{H}(r)=q(r)\, T_\text{H}=\left(1-{r_0^{d-2}\over r^{d-2}}\right)^{-\h} {(d-2)\over 4\pi r_0}\approx {(d-2)^\h\over 4\pi  r_0^\h (r-r_0)^\h}\,,
\ee
 where the final expression is the approximation appropriate for the Rindler region $r-r_0\ll r_0$.  The stress tensor of the Boulware vacuum is then
 \beq
T^\mu{}_{\nu}={\rm diag}\{ -\rho(r), p(r),\cdots, p(r) \}\,,
\label{mthree}
\eeq
with
\be
\rho(r)=-a T_\text{H}(r)^{d+1}\,,
\label{mthreeww}
\ee
and
\be
p(r)={1\over d} \rho(r)\,.
\label{zzthir}
\ee

\subsection{The stress tensor near an ECO}

Now we will argue that if we have an ECO at temperature $T_\text{ECO}=0$, then the stress tensor in the region just outside $R_\text{ECO}$ has the {\it same} value, to a first approximation, as the stress tensor (\ref{mthree})-(\ref{zzthir}) in the Boulware vacuum of the black hole. The key point will be that due to the compactness of the ECO, the  analogues of the wavemodes $f_{\{ l \},k}(r)$ in the ECO will have a large number of oscillations in the region $r>R_\text{ECO}$. These oscillations allow us to make local wavepackets out of these modes, and the local value of the stress tensor can be obtained to a good approximation from such wavepackets. Thus the detailed nature of the wavemode in the interior region $r<R_\text{ECO}$ becomes  irrelevant to the computation of the stress-energy outside $R_\text{ECO}$.

First consider the wavemodes $f_{\{ l \},k}(r)$ in the black hole metric.  These modes have the form
\be
f_{\{ l \},k}(r)={\chi_{\{ l \},k}(r)\over r^{d-1\over 2}}\,,
\ee
with $\chi_{\{ l \},k}(r)$ satisfying the equation
\be
- {d^2 \over d{r^*}^2}\chi_{\{ l \},k} +\left(1-{r_0^{d-2}\over r^{d-2}}\right)\left ({(d-1)^2\over 4} {r_0^{d-2}\over r^d}+ {L^2+{1\over 4}(d-1)(d-3)\over r^2} \right ) \chi_{\{ l \},k}=\omega^2 \chi_{\{ l \},k}\,,
\label{mfour}
\ee
where $\omega=|k|$ and $L^2=l(l+d-2), \, l=0, 1, 2, \dots$ is the value of the quadratic Casimir describing angular momentum. We have defined the `tortoise' coordinate $r^*$ through
$dr^*=\left(1-{r_0^{d-2}\over r^{d-2}}\right)^{-1} dr$, which in the near horizon region gives
\be
r^*\approx {r_0\over (d-2)} \ln \left ( {r-r_0\over r_0} \right ) \approx -{2r_0\over (d-2)} \log{r_0\over s}\,.
\label{zzfourt}
\ee
For later use we define the `effective potential' appearing in (\ref{mfour})
\be
V_{eff}\equiv \left(1-{r_0^{d-2}\over r^{d-2}}\right)\left ({(d-1)^2\over 4} {r_0^{d-2}\over r^d}+ {L^2+{1\over 4}(d-1)(d-3)\over r^2} \right )\,.
\ee

\begin{figure}
    \centering
    \includegraphics[scale=0.6]{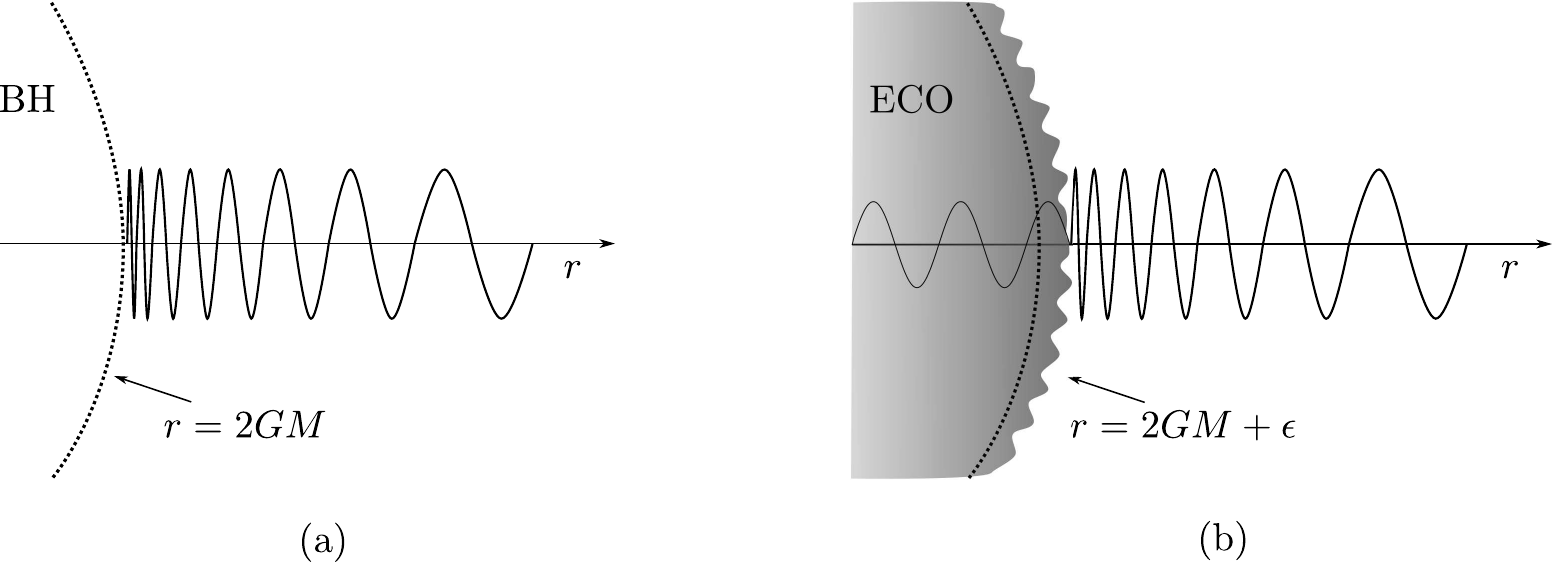}
    \caption{(a) In the black hole geometry, a wavemode oscillates an infinite number of times as it approaches the horizon. (b) In the ECO geometry, the wavemode oscillates a large number of times $n\gg 1$ before entering the ECO surface at $R_{\text{ECO}}$.}
    \label{ECOoscillations}
\end{figure}

In the near horizon region, $V_{eff}\r 0$, and we have
\be
 - {d^2\over d {r^*}^2}\chi_{\{ l \},k} =\omega^2 \chi_{\{ l \},k}\,,
 \label{freefield}
\ee
which gives
\be
\chi\sim e^{\pm i \omega r^*}\,.
\label{freefieldp}
\ee
Suppose we look at the region $r_0+\epsilon<r\lesssim 2r_0$, for $\epsilon\ll r_0$. The corresponding range for $r^*$ is
\be
{r_0\over (d-2)} \ln \left ( {\epsilon\over r_0} \right ) <r^*<0\,.
\ee
The phase of $\chi$ is seen to oscillate a number of times $n$ given by
\be
n\approx  {\omega\over 2\pi}{r_0\over (d-2)} \ln\left ( {r_0\over \epsilon} \right )\approx  {\omega\over 2\pi}{2r_0\over (d-2)} \ln \left ( {2\over (d-2)^\h }{r_0\over s} \right ) \,,
\label{ccone}
\ee
where in the last step we have written the coordinate interval $\epsilon$ in terms of the proper $s$ distance from the horizon. Note that $\omega r_0\sim 1$, since the energy of the typical quantum emitted is of order the black hole temperature $\sim 1/r_0$. For $\epsilon/r_0$ small (equivalently, $s/r_0$ small), we find that the number of oscillations near the horizon is $n\gg 1$.  As we approach the horizon, we have $\epsilon\r 0$ and the number of oscillations becomes infinite. We depict these oscillations in fig.\,\ref{ECOoscillations}(a).
 
Now consider the computation of the stress tensor for an ECO. We should first expand the field $\hat \phi$ in terms of field modes that satisfy $\square\hat \phi=0$ in the metric produced by the ECO:
\be
\hat\phi= \sum_{\{ l\},k} \Big (~   \hat b_{\{ l\},k}\, g_{\{ l \},k}(r) Y_{\{ l\}}(\{ \Phi\}) e^{-i\omega_{\{ l \},k}\, t}+\hat b^\dagger_{\{ l \},k}\,  g^*_{\{ l \},k}(r) Y^*_{\{ l \}}(\{ \Phi \} ) e^{i\omega_{\{ l \},k}\, t}  ~\Big )\,.
\label{elw}
\ee
Consider the situation where the ECO is at temperature $T_\text{ECO}=0$. Then the quantum field $\hat \phi$ will be in its lowest energy state $|\Psi_{\text{ECO},0}\rangle$ in the background geometry created by the ECO. This state is described by
\be
 \hat b_{\{ l\},k}|\Psi_{\text{ECO},0}\rangle=0\,.
 \ee
 A priori, the modes $g_{\{ l \},k}(r)$ will depend on the metric in the outside region  $r>R_\text{ECO}$ as well as the metric in the inside region $r<R_\text{ECO}$; in fact they are required to satisfy a smoothness condition at $r=0$ which is in the region $r<R_\text{ECO}$. But due to the compactness of the ECO, the modes  
  $g_{\{ l \},k}(r)$ will have a large number of oscillations $n$ in the region $R_\text{ECO}<r<2r_0$. Note that by condition ECO1, we have $s_{ECO}\ll s_c$, where $s_c$ is the compactness length scale $s_c$ defined  in (\ref{zzel}).   From (\ref{ccone}) we find, using  $r_0^{d-2}=\mu GM$, 
  \be
  n\gg {\omega\over 2\pi}{2r_0\over (d-2)} \ln \left ( {2\over (d-2)^\h }{r_0\over s_c} \right )\sim {\omega r_0}\ln \left ( {M\over m_p}\right )\sim \ln \left ( {M\over m_p}\right )\, ,
  \ee
  where in the last step we have again set $\omega r_0\sim 1$ since we have in mind a temperature for the ECO which is of order the black hole temperature. This large  number of oscillations is depicted in fig.\ref{ECOoscillations}(b). 
Due to this large number of oscillations of $g_{\{ l \},k}(r)$ in the region just outside $R_\text{ECO}$, any computation using these modes can be captured equally well by local wavepackets built from these modes. Thus we can compute the quantity $\langle \Psi_{\text{ECO},0}| \hat \phi(x)\,\hat\phi(y)|\Psi_{\text{ECO},0}\rangle$, 
 to a good approximation, by using the form of the modes $g_{\{ l \},k}(r)$ in the region $r>R_\text{ECO}$. If we assume that this region is just given by the Schwarzschild geometry (\ref{six}) with mass $M$, then the wavepackets made from modes $g_{\{ l \},k}(r)$ in the region $r>R_\text{ECO}$ will be approximately  the same as wavepackets made from modes $f_{\{ l \},k}(r)$ in the black hole geometry. In this approximation, we will get the same value for the stress tensor in the ECO as we had in the black hole. Thus we will have in the region $r>R_\text{ECO}$
\beq
T^\mu{}_{\nu}={\rm diag}\{ -\rho(r), p(r),\cdots, p(r) \}\,,
\label{mthreew}
\eeq
with
\be
\rho(r)=-a\, T_\text{H}(r)^{d+1}, ~~~~p={1\over d} \rho(r)\,.
\label{mthreewq}
\ee

\subsection{The ECO at a general temperature $T$}

We have seen above that the ECO at temperature $T_\text{ECO}=0$ has a negative energy density near its surface just like the negative energy density of the Boulware vacuum. This result was  based on the assumption that the metric in the region of interest was well approximated by the Schwarzschild metric. Let us continue with this approximation in mind and discuss the case of an ECO at a temperature $T_\text{ECO}\ne 0$. Later we will discuss how to take into account the departure of the metric at $r>R_\text{ECO}$ from the Schwarzschild form.

Thus suppose the ECO has a temperature at infinity given by $T_\text{ECO}$. At a radius $r$, the redshift results in a local temperature
\be
T_\text{ECO}(r)=q(r)T_\text{ECO}\,,
\ee
where $q(r)$ is given by (\ref{kone}). This radiation generates a stress tensor of the form (\ref{mthreew}) with
\be
\rho(r)=a\, T_\text{ECO}(r)^{d+1}, ~~~~p={1\over d} \rho(r)\,.
\ee

Taking into account the vacuum stress-energy (\ref{mthreew}),(\ref{mthreewq}) we find that,  with the above mentioned approximations,  the total stress tensor in the region just outside the ECO is of the form (\ref{mthreew}) with
\be
\rho(r)=a\, \left(T_\text{ECO}^{d+1}-T_\text{H}^{d+1}\right) q(r)^{d+1}, ~~~~p={1\over d} \rho(r)\,.
\label{vvone}
\ee

\section{Relating $S$  to $T$}\label{sec4}

Suppose that we were {\it given} that the temperature $T_\text{ECO}[M]$ of an ECO  had to be the same as the temperature of a black hole $T_\text{H}[M]$. Then, as we argue in this section, two conclusions follow. First, the entropy $S_\text{ECO}[M]$ of the ECO will have to agree with the black  hole entropy $S_{bek}[M]$. Second, the radiation rate from the ECO $\Gamma_\text{ECO}[M]$ will have to agree with the radiation rate $\Gamma_\text{H}[M]$ from the black hole. 

Having made these arguments,  we will be left with the task of showing that $T_\text{ECO}[M]=T_\text{H}[M]$, which will be the main task of this paper. This task will be tackled in the remaining sections. 

\subsection{Relating the entropy $S_{\text{ECO}}$ to the temperature}

Consider the entropy of any isolated body which has a large number of degrees of freedom. We have the relation
\be
TdS=dE\,.
\label{mfive}
\ee
Now suppose we are given  that our ECO has an temperature $T_\text{ECO}[M]$ that matched the Hawking temperature of the black hole, then we would have 
\be
S_\text{ECO}=\int T_\text{ECO}^{-1} {dM}=\int \left ( {4\pi r_0\over d-2}\right ) \left ( {(d-2) r_0^{d-3}\over G\mu} \right )  dr_0 = {\Omega_{d-1} r_0^{d-1}\over 4G}={A\over 4G}\,,
\label{meight}
\ee
where we have used (\ref{msix}) and (\ref{mseven}). 
The RHS of (\ref{meight}) is the Bekenstein Hawking entropy for the $d+1$ dimensional hole.

The above computation is not new; it is just the standard one through which the Bekenstein entropy of a black hole was obtained after the discovery that the hole radiates at  temperature $T_\text{H}$ \cite{bek,hawking}. This value of the black hole entropy was then reproduced through the Gibbons-Hawking computation of a Euclidean path integral \cite{gibbonshawking}. Here we are just noting that {\it any} isolated object with the same temperature function $T_\text{H}[M]$ as the black hole will have the same entropy as the black hole. The fact that the object is isolated tells us that we do not have to consider  additional terms in (\ref{mfive}) like $PdV$ or $\mu_i dN_i$ (here $\mu_i$ are chemical potentials and $N_i$ are the corresponding particle numbers).

\subsection{Relating the emission rate $\Gamma_{\text{ECO}}$ to the temperature}

We now address the radiation rate from the ECO. First consider a normal body, like a ball of steel. The rate of radiation from the ball cannot be determined if we just know the temperature $T$. For example, a ball painted black will radiate more energy than a ball painted white; this is because emission and absorption are related by detailed balance, and a black ball absorbs more readily than a white ball. Further, if the wavelength $\lambda$ of the emitted radiation is order $\lambda\gtrsim R$ where $R$ is the size of the body, then the radiation rate can also depend on the details of the shape of the body, and not just on its surface area.  

Now consider our ECO. Since the radius of the ECO is very close to the radius of the corresponding black hole, the surface area of the ECO will be almost the same as the surface area of the hole. But we have not assumed anything else about the nature of the surface of the ECO. Further, the wavelength $\lambda$ of the  radiation from a hole is $\lambda\sim r_0$, where $r_0$ is the radius of the hole. It may therefore appear that even if we are given that the ECO is at the same temperature $T_\text{H}$ as the corresponding black hole, we may not be able to say anything about the rate of radiation from the ECO.

But as we will now note, such is not the case. The radiation rate $\Gamma_\text{ECO}$ from the ECO  must equal the radiation rate $\Gamma_{\text{H}}$ to leading order.  We will first review the derivation of Hawking radiation from a black hole using Schwarzschild coordinates. Here we will note that the near horizon region is filled with a hot gas of quanta at the local Rindler temperature. These quanta tunnel through a high barrier at larger $r$ and reach infinity to give Hawking radiation. Next we will consider an ECO at the same temperature. The near surface region will be filled with quanta at the  same temperature as the corresponding black hole, and the barrier through which they must tunnel will also be approximately the same. Thus the radiation profile will agree between the black hole and the ECO. A crucial point in this argument  will be the compact nature of the ECO; this compactness leads to a separation between the near surface region which contains the hot gas of quanta, and the region further out which contains the potential barrier.

\begin{figure}
    \centering
    \includegraphics[scale=0.26]{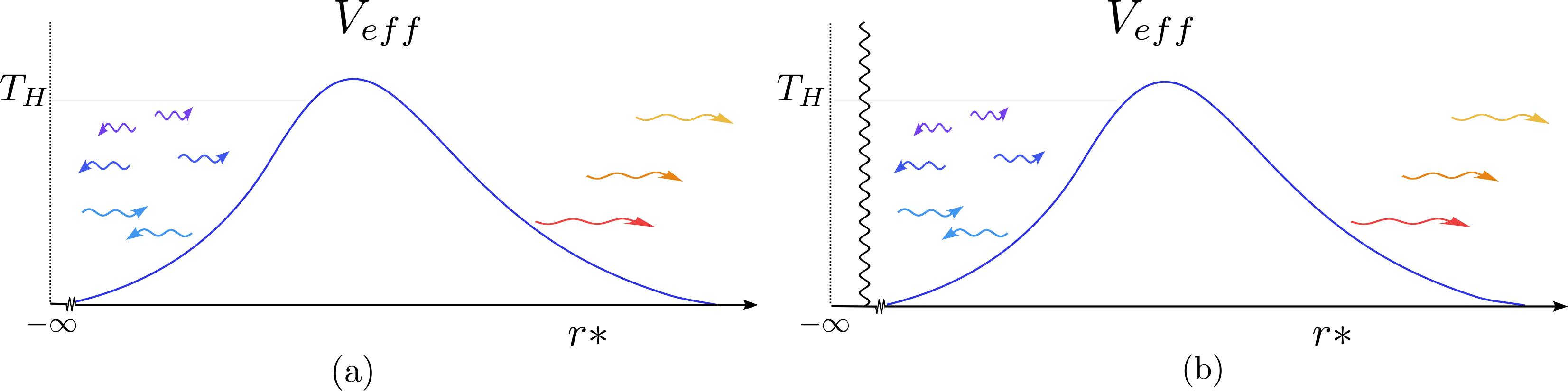}
    \caption{(a) In the black hole geometry, Rindler quanta are trapped near the horizon by the potential $V_{eff}$. Some low $l$ quanta manage to tunnel through this potential and escape to infinity as Hawking radiation. (b) In The ECO, we have a similar set of thermal quanta trapped near the ECO surface; these quanta must tunnel through essentially the same potential  to escape as radiation.}
    \label{ECOscalarpotential}
\end{figure}

For radiation from the black hole, we consider the wave-equation (\ref{mfour}) for modes in the black hole geometry. We look at three regions: (i) close to the horizon (ii) further out, in the region of the potential barrier (iii) the region near infinity. The setup is pictured in  fig.\,\ref{ECOscalarpotential}(a).

\begin{enumerate}[label=(\roman*)]
\item  Close to the horizon, the effective potential $V_{eff}$ vanishes due to the factor $1-{r_0^{d-2}\over r^{d-2}}$, and the wave-equation becomes
\be
- {d^2 \over d{r^*}^2}\chi_{\{ l \},k} =\omega^2 \chi_{\{ l \},k}\,.
\label{mfoure}
\ee
In this near-horizon region we can describe the geometry by the Rindler coordinates (\ref{knine}), and (\ref{mfoure}) yields the Rindler modes for our field $\hat\phi$. In this spacetime region it is useful to think in terms of the proper distance $s$ from the horizon radius $r_0$. The spacetime is filled with a thermal gas of Rindler excitations at temperature $T_\text{H}(s)={1\over 2\pi s}$; this temperature is just the appropriately blueshifted value of the Hawking temperature $T_\text{H}$ at infinity. The radiated quanta have energy of order the Hawking temperature; thus we have $\omega \sim T_\text{H} \sim 1/r_0$ in (\ref{mfoure}).

\item  At larger $r$, the potential $V_{eff}$ becomes important, and gives a high barrier for the modes $\chi_{\{ l \},k}$. The reason that the barrier is high is that the angular momentum $L$ is large for a typical mode in the thermal gas near $r=r_0$. At a proper distance $s$ from the horizon radius $r_0$, the wavelength of the typical quantum is $\lambda\sim s$, and thus the typical value of $L$ in the thermal gas  is
\be
L\sim {r_0\over s}\,.
\ee
The barrier height for $s\ll r_0$ is (transforming coordinates from $r$ to $s$)
\be
V_{eff}\approx {s^2L^2\over r_0^4}\,.
\ee
Thus at distances $s\sim r_0/L$,  the potential term in (\ref{mfour}) becomes of order  $1/r_0^2\sim \omega^2$. At larger $s$, the potential rises to a value higher than the energy term $\omega^2$ on the RHS of (\ref{mfour}), and at $s\sim r_0$, we get 
\be
V_{eff}\sim  {L^2\over r_0^2}\sim L^2\omega^2\,.
\ee
We see that for $L\gg1 $, the height of the barrier is much larger than the energy term $\omega^2$ in the wave-equation (\ref{mfour}).

\item At $r\r \infty$, we again have $V_{eff}\r 0$, and we get freely propagating waves satisfying (\ref{mfoure}). From the discussion in (ii) above, the potential $V_{eff}$ confines the the large $L$ angular  harmonics to a small distance $s\sim r_0/L$ from the horizon, and only models with $L\sim 1$ tunnel through the barrier and escape to infinity.

\end{enumerate}

Thus we can derive the rate of Hawking radiation from a black hole as follows. Consider any spherical harmonic $Y_{\{ l\}}(\{ \Phi\})$. The corresponding radial function $\chi_{\{ l \},k}$ gives  a freely traveling outward wave at $r^*\r -\infty$. This wave encounters the potential barrier $V_{eff}$, whereupon a part of the wave gets reflected back and a part tunnels out and escapes to infinity as `Hawking radiation'. The large $L$  modes get reflected back strongly; thus only modes with  $L\sim 1$  escape to infinity. Using the potential $V_{eff}$ we can compute the probability $P[\{l\},\omega]$ for the wavemode to tunnel from the region $r^*\r-\infty$ out to $r^*\r\infty$. 

We assume an occupation number of the modes $\chi_{\{ l \},k}$ at $r^*\r -\infty$ given by the local Rindler temperature $T_\text{H}(s)=1/(2\pi s)$. This occupation number determines the flux of such modes that is incident on the potential barrier from the side $r^*\r-\infty$. Multiplying this flux by the probability of tunneling $P[\{l\},\omega]$ gives the radiation rate $\Gamma_\text{H}[\{ l\}, \omega]$.

Note that in this derivation of the Hawking radiation rate $\Gamma_\text{H}[\{ l\}, \omega]$ we have used only the exterior of the hole $r>r_0$. Hawking's original derivation had considered modes that straddle the horizon; these modes get distorted to produce  a pair of quanta, one inside the hole and one escaping to infinity as  radiation. In the derivation above, the smooth horizon that Hawking assumed is taken into account by the assumption of the  thermal Rindler state just outside the horizon; this thermal state is equivalent to the local Minkowski vacuum at the horizon in Kruskal coordinates.

 \b

Now consider the radiation from the ECO. We depict the situation in fig.\,\ref{ECOscalarpotential}(b). In the black hole geometry (\ref{six}), the coordinate $r^*$ goes to $-\infty$ as $r\r r_0^+$ (eq.\,(\ref{zzfourt})). In the ECO, the Schwarzschild geometry gets completely modified at $r<R_\text{ECO}$, but note that $R_\text{ECO}-r_0$ is small. By condition ECO3, the Schwarzschild  geometry is obtained to a good approximation for points outside the compactness scale $s_c$. From (\ref{zzfourt}), this region is   
\be
r^*\gtrsim  -r_0\ln {r_0\over s_c}\,.
\ee
In the region 
\be
-r_0\ln {r_0\over s_c}\lesssim r^* \ll r_0\,,
\ee
the effective potential $V_{eff}$ is small due to the vanishing of the factor $(1-{r_0^{d-2}\over r^{d-2}})$, and we get the free-field behavior (\ref{freefield}),(\ref{freefieldp}) for $\chi_{\{ l \},k}$ in (\ref{mfour}). We have a gas of quanta in this region at a local temperature 
\be
T_{\text{ECO}}(s)=q(s)\, T_\text{ECO} \approx {T_\text{ECO}\over T_\text{H}} {1\over 2\pi s}\,,
\label{mten}
\ee
where $q(s)$ is given by (\ref{kone}). The potential $V_{eff}$ in the region $r^*\gtrsim -r_0\ln {r_0\over s_c}$  is, by condition ECO3, approximately  the same as in the case of the black hole, so the tunneling probability for a quantum with a given $L, \omega$ will be the same, to leading order,  for the black hole and the ECO.

Thus we can compute the radiation from the ECO in a manner that parallels the computation of radiation from the black hole. In the ECO geometry, we have a thermal distribution of quanta near $r=R_\text{ECO}$ with the temperature (\ref{mten}). Quanta with large $L$ reflect off the barrier and return to the surface of the ECO, where they thermalize with the degrees of freedom in the ECO and get reemitted to the region $r>R_\text{ECO}$ (with in general a different value of $L$). The quanta with low $L$ which escape to infinity determine the radiation rate
$\Gamma_\text{ECO}[\{ l\}, \omega]$. 

The probability of tunneling for these low $L$ quanta is the same, to leading order, as the probability of tunneling in the black hole geometry, since for low $L$ the potential barrier is significant only in region $r-r_0\sim r_0$, which is far from the ECO surface at $r=R_\text{ECO}$. 

If $T_\text{ECO}=T_\text{H}$, then the flux incident on the barrier from $r^*\r-\infty$ is the same, to leading order, for the ECO and the black hole. Thus we will get 
\be
\Gamma_\text{ECO}[\{ l\}, \omega]\approx \Gamma_{H}[\{ l\}, \omega]\,.
\ee

\b

Let us recall how the conditions ECO1-ECO3 were important in obtaining the above conclusion. 
First consider the condition ECO1. As we have seen, only modes with $L\sim 1$ are radiated in an appreciable manner. The effective potential $V_{eff}(r)$ for such modes is peaked around $(r-r_0)/r_0\sim 1$, and vanishes for $(r-r_0)/r_0\ll 1$ due to the factor $1-{r_0^{d-2}\over r^{d-2}}$. If $(R_\text{ECO}-r_0)/r_0$ is not very small, then the radiated modes in the ECO pass through only that part of the potential which is in the region $r>R_\text{ECO}$ in contrast to modes in the black hole which tunnel through the full range $r_0<r<\infty$. For $(R_\text{ECO}-r_0)/r_0\ll1$, the modes with $L\sim 1$ encounter essentially the same potential in the ECO and in the black hole.

The condition ECO2 of high redshift at the surface of the ECO  is important because it implies a high local temperature for quanta near $r=R_\text{ECO}$. This high temperature translates to a short wavelength, which gives a well defined local thermal distribution of quanta in the region $(r-R_\text{ECO})/r_0\ll 1$. Having such a thermal distribution allows us to separate the radiation computation into two parts. One part is the computation of a tunneling probability through the barrier (which peaks at $r-r_0\sim r_0$) and the other part is the computation of the number density of quanta present near $r=r_\text{ECO}$. If we did not have such a separation, then we would not be able to assume a black body distribution of quanta at some local temperature $T_{\text{ECO}}(r)$; the potential $V_{eff}$ would become relevant even for finding the distribution of modes near $r=R_\text{ECO}$.

Finally, the condition ECO3 allowed us to use the effective potential $V_{eff}$ in the region from $s\sim  s_c$ out to infinity.

\section{A heuristic argument for the relation $T_{\text{ECO}}\approx T_\text{H}$}\label{sec5}

In section \ref{sec4} we have seen that \textit{if} we are given that $T_{\text{ECO}} =T_\text{H}$, then the entropy and radiation rates of the ECO will agree with the corresponding quantities for the black hole. We now pass on to our main task: arguing that an ECO that satisfies our conditions ECO1-ECO3 cannot have an arbitrary temperature, but rather must have $T_{\text{ECO}} \approx T_\text{H}$. It will be clear from our discussion that the approximation in this relation will become better as we take the distance $s_{\text{ECO}}$ to be smaller.

In this section, we will make our first pass at arguing for this equality of temperatures, in the process obtaining the compactness condition (\ref{sevenq}). In this first pass, we will not be completely consistent in our approximations, in the following sense.  We will need to use the energy density of the radiation near the surface of the ECO. This energy density at a radius $r$ depends on the value of the redshift at $r$. This redshift, in turn, is affected by the energy density of the radiation itself. But in the analysis of this section, we will ignore this feedback of the radiation on the metric, assuming instead the redshift implied by the usual black hole metric in the region $r > R_{\text{ECO}}$. In the next section, we will remedy this inaccuracy by solving the Tolman-Oppenheimer-Volkoff equation in the region near $r = R_{\text{ECO}}$. 

\subsection{The energy of radiation near the ECO}

 Consider the near-surface region  outside $R_{\text{ECO}}$, described by the condition  $r-R_{\text {ECO}}\ll r_0$. As we saw in section\,\ref{sec4}, the effective potential $V_{eff}$ traps  the radiation from the ECO in this near-surface region, so that we have a   thermal gas of quanta at some local temperature $T_{\text{ECO}}(r)$ which depends on $r$. Far from the ECO we just have outgoing radiation in low $l$ harmonics, so the energy density is very low. Thus we will take the thermal distribution of the near-surface region to be truncated so that it is nonvanishing only in the  the region  $r\lesssim 2r_0$; in fact as we will see the energy density is appreciable only very close to $r=R_{\text{ECO}}$. 

 This radiation in the region $R_{\text{ECO}} < r < 2r_0$ has a total energy which we will call $E_{\text{rad}}$. Note that due to the negative vacuum energy in the region outside the ECO, $E_{\text{rad}}$ will be positive if $T_{\text{ECO}} > T_\text{H}$ and $E_{\text{rad}}$ will be negative if $T_{\text{ECO}} < T_\text{H}$.

Now consider the mass function $M(r)$ defined in (\ref{ktwop}). Since the mass at infinity is $M$, we will have
\begin{equation}
    M(R_{\text{ECO}}) = M - E_{\text{rad}}\,.
    \label{eq.Mrad}
\end{equation}
The quantity $M(R_{\text{ECO}})$ can be thought of as the mass contained inside the `core' of the ECO -- the region which contains any nontrivial quantum gravitational dynamics. Recall that the horizon radius $r_0$ of a black hole is related to its mass $M$ by the relation (\ref{mseven}). Correspondingly, we define a radius $\tilde{r}_0$ through
\begin{equation}
    \tilde{r}_0^{d-2} = \mu G M(R_{\text{ECO}})\,,
\end{equation}
where the radius $\tilde{r}_0$ would be the radius of a black hole with mass $M(R_{\text{ECO}})$ if we had a Schwarzschild  geometry with  mass $M(R_{\text{ECO}})$. The equation \eqref{eq.ECO4} requires
\begin{equation}
    1 - \frac{\mu G M(R_{\text{ECO}})}{R_{\text{ECO}}^{d-2}} > 0\,,
\end{equation}
so that 
\begin{equation}
    \tilde{r}_0 < R_{\text{ECO}}\,.
\end{equation} 
Thus $\tilde{r}_0$ is a radius that is inside a region that we will not directly address.
Nevertheless, the coordinate separation $R_{\text{ECO}}-\t r_0$ will play an important role in the discussion below. We will generally find that
${\t r_0-r_0}\ll r_0$
and so in many steps below we will use the approximation
\be
\t r_0\approx r_0\,,
\ee
to simplify our expressions. 

Consider the geometry generated by a mass $M(R_\text{ECO})$ confined within the radius $r=R_{\text{ECO}}$, and no energy density at $r>R_{\text{ECO}}$. For $r>R_{\text{ECO}}$ the metric for this geometry is of the Schwarzschild form
\begin{align}
ds^2&= -\Big ( 1-\left({\t r_0\over r}\right)^{d-2}\Big ) dt^2 + {dr^2\over 1-\left({\t r_0\over r}\right)^{d-2}.}+r^2d\Omega_{d-1}^2 \, .
\label{sixm}
\end{align}
As noted above, in the actual ECO, there is a nonzero energy density $\rho$ in the region $r>R_{\text{ECO}}$. But we will be ignoring the deformation of the metric due to this  $\rho$ in some of steps below; in these steps we will refer to the metric (\ref{sixm}).

With the  metric (\ref{sixm}), the redshift at a radius $r$ is given by
\be
\t  q(r)=(-g_{tt}(r))^{-\h} \approx {\t r_0^\h \over (d-2)^\h (r-\t r_0)^\h}\approx {r_0^\h \over (d-2)^\h (r-\t r_0)^\h}\,.
\label{koneqm}
\ee
Here we have added a tilde to the variable $q$ to denote the fact that this redshift $\t q$ corresponds to the redshift in the geometry with horizon radius $\t r_0$.    
The energy density at radius $r$ is then 
\be
\rho(r)=a \left(T_\text{ECO}^{d+1}-T_{\text{H}}^{d+1}\right)\,  \t q(r)^{d+1}\, .
\label{xxten}
\ee
The total energy of this radiation in the region $R_\text{ECO}<r<\infty$ is
\bea
E_{\text{rad}}&\approx&\int_{r=R_{\text{ECO}}}^\Lambda dr \,\Omega_{d-1}\, r^{d-1}\, \rho(r)\, , \nn
&\approx &a \left(T_\text{ECO}^{d+1}-T_{\text{H}}^{d+1}\right)\,\Omega_{d-1} \int_{r=R_{\text{ECO}}}^\Lambda d r\, r^{d-1}\,\t q( r)^{d+1}\, ,\nn
 &\approx & a \left(T_\text{ECO}^{d+1}-T_{\text{H}}^{d+1}\right)\,\Omega_{d-1} \int_{r=R_{\text{ECO}}}^\Lambda d r \,r^{d-1}_0\left({r_0\over (d-2)(r-\t r_0)}\right)^{d+1\over{2}}\, , \nn
 &\approx&a \left(T_\text{ECO}^{d+1}-T_{\text{H}}^{d+1}\right){2\,\Omega_{d-1} \over (d-2)^{d+1\over 2}(d-1)}{r_0^{3d-1\over 2}\over  (R_\text{ECO}-\t r_0)^{d-1\over 2} }\, ,
 \label{ccten}
\eea
where $\Lambda=2r_0$ is  the cutoff we had chosen for our thermal bath and where we are ignoring the energy of the outgoing modes at $r\gtrsim 2r_0$. 

We will also be interested in the energy of a thin shell outside $r=R_\text{ECO}$. Note that the energy density (\ref{xxten}) falls off as a power of $r-\t r_0$. We  write
\be
\Delta r \equiv R_\text{ECO}-\t r_0\, .
\ee
and  define our thin shell as
\be
R_\text{ECO}<r< R_\text{ECO}+\Delta r.
\label{vthree}
\ee
The energy of such a shell will be
\be
E_\text{rad}^{\text{shell}}=\left(1-{1\over 2^{d-1\over 2}}\right) E_\text{rad}\,.
\label{vfive}
\ee
It will be helpful to write
\be
T_\text{ECO}=\eta_\text{T} \,T_{\text{H}}\, .
\label{vtwo}
\ee
Noting the expression (\ref{msix}) for $T_{\text{H}}$, we then get
\bea
E_\text{rad}&=&a\left(\eta_\text{T}^{d+1}-1\right) {2(d-2)^{d+1\over 2}\Omega_{d-1} \over (4\pi)^{d+1}(d-1)}{r_0^{d-3\over 2}\over  \left(R_\text{ECO}-\t r_0\right)^{d-1\over 2} }\, ,\nn
&\equiv&  {C_1\,r_0^{d-3\over 2}\over \left (R_\text{ECO}-\t r_0\right)^{d-1\over 2} }\,\label{eq.Erad} ,
\eea
where $C_1$ is a dimensionless constant of order unity. Note that $C_1>0$ for $T_\text{ECO}>T_\text{H}$, and $C_1<0$ for $T_\text{ECO}<T_\text{H}$. We will have $C_1=0$ if $T_\text{ECO}=T_\text{H}$; i.e., $\eta_\text{T}=1$. Our goal will be to show that this is the only allowed value for $\eta_\text{T}$ for an ECO.

\subsection{An outline of the argument}\label{secoutline}

\begin{figure}
    \centering
    \includegraphics[scale=0.7]{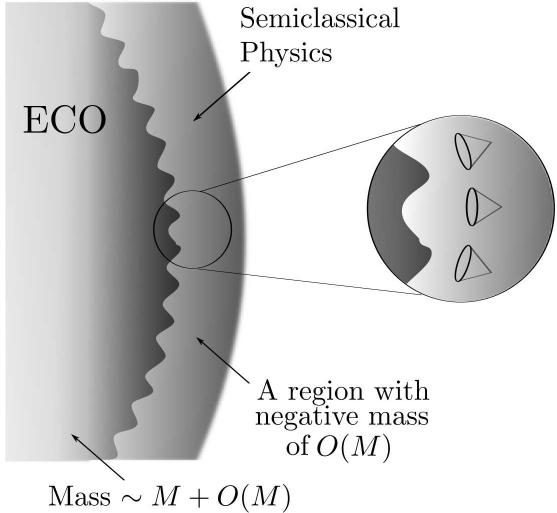}
    \caption{The argument for an ECO with $T_{\text{ECO}}<T_\text{H}$ and a  surface $r=R_{\text{ECO}}$  that is just a planck length outside the horizon radius. The region just outside $R_{\text{ECO}}$ has a large negative energy density due to the negative vacuum energy. Thus the core of the ECO (the region depicted with a jagged boundary) must have a mass significantly more than $M$. Since the radius of this core  is very close to the horizon radius $r_0$ for mass $M$, this core must be inside its own horizon; this fact is depicted in the magnified region by light cones that point inwards. Thus such an ECO cannot exist as a time-independent configuration.}
    \label{ecofinalNM}
\end{figure}

Let us first sketch the nature of the argument, before moving onto more detailed estimates. Let the spacetime be 3+1 dimensional for simplicity. We consider the cases $T_{\text{ECO}}<T_{H}$ and  $T_{\text{ECO}}>T_{H}$ in turn.

\subsubsection{The case  $T_{\text{ECO}}<T_{H}$}

Let us start with a simple case, depicted in fig.\,\ref{ecofinalNM}. Suppose $T_\text{ECO}=0$. Suppose further that the ECO surface, which is at $r=R_{\text{ECO}}$, is only a planck distance outside the horizon radius $r_0=2GM$; i.e., $s_{ECO}= l_p$ which gives $(R_\text{ECO}-\t r_0)\sim l_p^2/r_0$.  The vacuum energy density outside $R_\text{ECO}$ is negative and order planck scale just outside the surface $r=R_\text{ECO}$. We find from (\ref{eq.Erad})
\be
E_\text{rad}\sim -{1\over (R_\text{ECO}-\t r_0)}\sim -{r_0\over l_p^2}\sim -M\,,
\ee
where we have used the fact that in 3+1 dimensions, $G=l_p^2$. Thus the vacuum energy outside the core of the ECO is negative and of the same order as the mass $M$ seen at infinity. 

For concreteness, let us assume that $E_\text{rad}=-\h  M$. Then from (\ref{eq.Mrad}), we have
\be
M(R_\text{ECO})=M-E_\text{rad}={3\over 2}M\, .
\label{vone}
\ee

Now consider the validity of the relation (\ref{eq.ECO4}) at $r=R_\text{ECO}$.   Recall that  $R_\text{ECO}\approx 2GM$.  Thus we have
\be
1-{2GM(R_\text{ECO})\over R_\text{ECO}}\approx 1-{3GM\over 2GM}=-\h <0\, .
\ee
This contradicts the requirement \eqref{eq.ECO4}, which was required for regularity of the geometry. 

Put another way, the vacuum energy outside the ECO contributes a negative value $-\h M$ to the overall mass, which means the mass of the core has to be ${3\over 2}M$ to yield a total  mass $M$ at infinity. The Schwarzschild radius for this mass ${3\over 2}M$ is $3GM$, so that the core of the ECO is deep inside its own horizon radius. Recall that in classical general relativity, any particle inside a horizon must move towards smaller values of $r$ by the inward pointing structure of light cones inside the horizon. Noting that semiclassical dynamics was required to hold at $r\ge R_\text{ECO}$, we conclude that the particles at the surface $r=R_\text{ECO}$ cannot stay at a fixed radius $R_\text{ECO}$; instead, the core of the ECO must collapse. Thus we conclude that we cannot have such an ECO with $T_\text{ECO}=0$.

The argument does not change in any significant way   if we take some other temperature $T_\text{ECO}<T_\text{H}$. Suppose we take $T_\text{ECO}=\h\, T_{\text{H}}$. Then $\eta_\text{T} =\h$ in (\ref{vtwo}), as compared to the value $\eta_\text{T}=0$ for the case $T_\text{ECO}=0$. Using (\ref{eq.Erad}), we see that the energy $E_\text{rad}(\eta_\text{T}=\h)$ is smaller than  $E_\text{rad}(\eta_\text{T}=0)$  by a factor 
\be
\frac{E_\text{rad}(\eta_\text{T}=\h)}{E_\text{rad}(\eta_\text{T}=0)}={1-\left(\h\right)^{4}\over 1}={15\over 16}\,,
\ee
but this does not affect the nature of the argument we had outlined above for the case  $T_\text{ECO}=0$.

\begin{figure}
    \centering
    \includegraphics[scale=0.7]{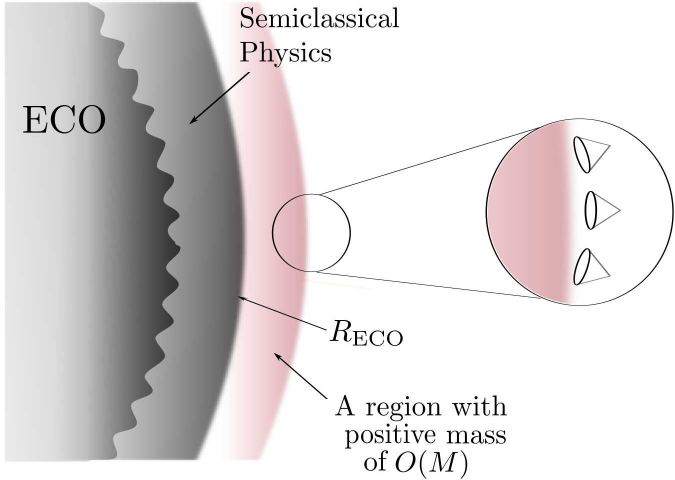}
    \caption{The argument for an ECO with $T_{\text{ECO}}>T_\text{H}$, and an surface at $R_{\text{ECO}}$ where the redshift is very high. This high redshift implies that $R_{\text{ECO}}$ is a very small distance outside the horizon radius $\t r_0$ corresponding to the mass inside $R_{\text{ECO}}$.  The region just outside $R_{\text{ECO}}$ has a large positive energy density due to the thermal radiation. A thin shell of this radiation, depicted as the outer band, has a significant amount of mass. Then the total mass in the region inside the outer boundary $r_{outer}$ of this shell is such that the corresponding horizon radius $r_{0,outer}$  is larger than $r_{outer}$. Thus we again find that such an ECO cannot exist as a time-independent configuration.}
    \label{ecofinalPM}
\end{figure}

\subsubsection{The case  $T_{\text{ECO}}>T_{H}$}

Now consider the case $T_\text{ECO}>T_\text{H}$, which is depicted in fig.\,\ref{ecofinalPM}. Condition ECO2 requires a high redshift at the surface $R_{ECO}$, which leads to a high local temperature  for the radiation. Let us assume for our example that this temperature is planck scale.  Then the energy density at the ECO surface is also planck scale. The energy of radiation outside $R_{\text{ECO}}$ is $E_\text{rad}\sim M$. 

For concreteness, let us assume that $E_\text{rad}=\h M$.  Then the mass of the core of the ECO is $M(R_\text{ECO})=\h M$. Note that the horizon radius corresponding to this mass $M(R_\text{ECO})$ is $\t r_0=GM$.  Now we observe that in order to get the required high redshift at $R_{ECO}$, the mass inside $R_{\text{ECO}}$ had to be compact enough to generate this high redshift. More precisely, we need the surface $R_{\text{ECO}}$ to be $\sim l_p$ outside the radius $\t r_0=GM$; thus we write $R_{\text{ECO}}\approx GM$. 

Now consider a shell of  radiation, with width $ l_p$, just outside the ECO surface. Thus the outer boundary of this shell $r_{outer}$ is very close to $R_{\text{ECO}}$, which was very close to $\t r_0=GM$.  Thus $r_{outer}\approx GM$. 

By (\ref{vfive}), the energy of the shell is $E_{rad}^{shell} = \h E_{rad}=0.25M$.  Thus the total mass inside the radius $r_{outer}$ is  is $M(r_{outer})=0.5 M + 0.25 M=0.75M$. Now consider the validity of the relation (\ref{eq.ECO4}) at $r=R_\text{ECO}$. We have
\be
1-{2GM(r_{outer})\over r_{outer}}\approx 1-{1.5GM\over GM}=-0.5 <0\,,
\ee
which contradicts the requirement (\ref{eq.ECO4}).  

In other words, the core of the ECO had to be very compact to yield a high redshift at $R_{\text{ECO}}$. We then find that a thin shell just outside this core adds enough mass so that  the system given by core+shell is inside its own horizon radius; thus such an ECO cannot exist.

\subsection{Defining a useful  scale $\Delta r_{crit}$}

Let us review the quantities which played a role in the above arguments. The difference  $r_0-\t r_0$ described the difference between the Schwarzschild radius $r_0$ for the mass $M$ of the ECO, and the the Schwarzschild radius $\t r_0$ of the core of the ECO. This difference  stems from the fact that a part $E_\text{rad}$ of the mass $M$ of the ECO is carried by the radiation surrounding the core of the ECO. The energy $E_\text{rad}$ in turn depends on 
\be
\Delta r \equiv (R_\text{ECO}-\t r_0)\,.
\label{zzninet}
\ee
 A very small value of $(R_\text{ECO}-\t r_0)$ gives a very large $E_\text{rad}$ and thus a very large $r_0-\t r_0$. Conversely, a very large $(R_\text{ECO}-\t r_0)$ gives a very small $E_\text{rad}$ and thus a very small $r_0-\t r_0$. 

Given the above, it will be useful to define a critical value $\Delta r_{crit}$ for $R_\text{ECO}-\t r_0$, such the corresponding radiation energy $E_\text{rad}$ shifts the horizon radius by an amount which is again $r_0-\t r_0=\Delta r_{crit}$. Recalling that the horizon radius satisfies $r_0^{d-2}=\mu G M$, we note that small shifts of this radius are given by $\delta r_0\approx {\mu G \delta M\over (d-2) r_0^{d-3}}$. Setting $\delta M=-E_\text{rad}$, we find that
\be
|r_0-\t r_0| \approx  {\mu G |E_\text{rad}|\over (d-2) r_0^{d-3}}\approx { {\mu G |C_1| \over  (\Delta r)^{d-1\over 2} (d-2) r_0^{d-3\over 2}}}\, ,
\label{zztwenty}
\ee
where we have used the expression for $E_{\text{rad}}$ from \eqref{eq.Erad}.
We define the critical separation $\Delta r_{crit}$ as the value of $\Delta r$ when this shift  $|r_0-\t r_0|$ equals $\Delta r$ defined in (\ref{zzninet}). 
Thus we have
\be \label{eq.rcrit}
\Delta r_{crit} ={ {\mu G |C_1| \over  (\Delta r_{crit})^{d-1\over 2} (d-2)\, r_0^{d-3\over 2}}}\, .
\ee
which gives
\be
\Delta r_{crit} = \left ( { {\mu G |C_1| \over  (d-2)\, r_0^{d-3\over 2}}} \right )^{2\over d+1}\, .
\ee
To get an idea of the scale  $\Delta r_{crit}$, we write it in terms of a proper distance $s_{crit}$ using the relation (\ref{knine}) between coordinate distance from the horizon and proper distance  from the horizon. Then we get
\be
s_{crit}\sim \left ( {M\over m_p}\right )^{2\over (d-2)(d+1)} l_p\sim s_c\,,
\label{zzttwo}
\ee
where the compactness scale $s_c$ was defined in (\ref{zzel}). 

\subsection{The argument for $T_\text{ECO}<T_\text{H}$}

Let us assume that $T_\text{ECO}<T_\text{H}$, so that $\eta_\text{T}<1$. As noted above, a crucial role in the argument is played by the quantity $\Delta r = R_\text{ECO}-\t r_0$. We will consider separately the following two possibilities:

\begin{enumerate}[label=(\alph*)]

\item $\Delta r \ll \Delta r_{crit}$\,:

Since $\eta_{\text{T}}<1$,  we have $E_\text{rad}<0$. Further,
\be
|E_\text{rad}|= {|C_1|\, r_0^{d-3\over 2}\over  (R_\text{ECO}-\t r_0)^{d-1\over 2}}\gg {|C_1| \,r_0^{d-3\over 2}\over  (\Delta r_{crit})^{d-1\over 2}}\, .
 \ee
Since $E_\text{rad}<0$, we have $\t r_0>r_0$. From (\ref{zztwenty}) we have 
 \be
 (\t r_0-r_0)=  {\mu G |E_\text{rad}|\over (d-2)\, r_0^{d-3}}\gg  { {\mu G |C_1| \over  (\Delta r_{crit})^{d-1\over 2} (d-2)\, r_0^{d-3\over 2}}}\sim \Delta r_{crit}\,,
 \label{zztone}
 \ee
 where the last relation follows from using the definition (\ref{eq.rcrit}) of $\Delta r_{crit}$. 
 
 Now, since $R_\text{ECO}>\t r_0$, we find that
 \be
 R_\text{ECO}-r_0=(R_\text{ECO}-\t r_0)+(\t r_0-r_0)>(\t r_0-r_0)\gg \Delta r_{crit} \,,
 \ee
 where in the last step we have used (\ref{zztone}). 
 Given that the separation $\Delta r_{crit}$ corresponds to the proper distance scale $s_c$ by (\ref{zzttwo}), we see that our ECO is not sufficiently compact to satisfy condition ECO1.

 \item $\Delta r \gtrsim \Delta r_{crit}$\, :
 
 In this situation the redshift at the ECO surface is, using (\ref{kone})
   \be
 q(R_\text{ECO})\approx  {r_0^\h\over (d-2)^\h (R_\text{ECO}-\t r_0)^\h}\lesssim {r_0^\h\over  (\Delta r_{crit})^\h}\sim \left ( {r_0\over l_p}\right)^{d-1\over d+1}\, .
 \label{ccthree}
 \ee
 This violates the condition ECO2  (eq.\,(\ref{ksixq})) which requires that we have a higher redshift than (\ref{ccthree}) at $R_\text{ECO}$.
 
 Thus we find that for $T_\text{ECO}<T_\text{H}$, we cannot get an ECO satisfying the conditions ECO1-ECO3.
 
 \end{enumerate}
 
 \subsection{The estimate for $T_\text{ECO}>T_\text{H}$}
 
 Now we consider the case $T_\text{ECO}>T_\text{H}$, so that $\eta_\text{T}>1$. Again we proceed to examine the two possibilities:
 
 \begin{enumerate}[label=(\alph*)]
     \item$\Delta r \ll \Delta r_{crit}$\, :
     
     Since $\eta_T>1$,  we have $E_\text{rad}>0$. Further,
 \be
E_\text{rad}= {C_1\, r_0^{d-3\over 2}\over  (R_\text{ECO}-\t r_0)^{d-1\over 2}}\gg {C_1\, r_0^{d-3\over 2}\over  (\Delta r_{crit})^{d-1\over 2}}\, .
 \ee
 Since $E_\text{rad}>0$, we have $r_0>\t r_0$, and 
 \be
 ( r_0-\t r_0)=  {\mu G E_\text{rad}\over (d-2) r_0^{d-3}}\gg   { {\mu G |C_1| \over  (\Delta r_{crit})^{d-1\over 2} (d-2)\, r_0^{d-3\over 2}}}\sim \Delta r_{crit}\, .
 \ee
Now consider the thin shell outside $r=R_\text{ECO}$ with coordinate width $\Delta r$ as defined in (\ref{vthree}). We have from (\ref{vfive})
\be
E_\text{rad}^{\text{shell}}= \left(1-{1\over 2^{d-1\over 2}}\right){ C_1\, r_0^{d-3\over 2}\over  (R_\text{ECO}-\t r_0)^{d-1\over 2}}\gg {C_1\, r_0^{d-3\over 2}\over  (\Delta r_{crit})^{d-1\over 2}}\, .
 \ee
 Let us define $\t r_{0, outer}$ as the Schwarzschild radius corresponding to the mass within radius $r_{outer}$; i.e., we define $r_{0, outer}$ through the condition $r_{0,outer}^{d-2}\equiv \mu G M(r_{outer})$. Then we have
\be
r_{0,outer}-\t r_0 \approx {\mu G E^{\text{shell}}_\text{rad}\over (d-2) r_0^{d-3}}\gg    { {\mu G |C_1| \over  (\Delta r_{crit})^{d-1\over 2} (d-2)\, r_0^{d-3\over 2}}}\sim \Delta r_{crit}\, ,
\label{zztthree}
\ee
where in the last step we have used (\ref{eq.rcrit}). 
Now recall that we had chosen the coordinate width of the shell as described in (\ref{vthree}), which says
\be
r_{outer}-R_{\text{ECO}}= R_\text{ECO}-\t r_0 \,.
\ee
Thus since $\Delta r =(R_\text{ECO}-\t r_0)\ll \Delta r_{crit}$, we find
\be
r_{outer}-R_{\text{ECO}}\ll  \Delta r_{crit}\,.
\label{xxone}
\ee
On the other hand, from (\ref{zztthree}) we have
\be
r_{0,outer}-\t r_0\gg  \Delta r_{crit}\,.
\label{xxtwo}
\ee
Subtracting (\ref{xxone}) from (\ref{xxtwo}) we get
\be
(r_{0,outer}-r_{outer})+(R_{\text{ECO}}-\t r_0)\gg  \Delta r_{crit}\,.
\ee
Again using $(R_\text{ECO}-\t r_0)\ll \Delta r_{crit}$ we find
\be
(r_{0,outer}-r_{outer})\gg  \Delta r_{crit}\,.
\ee
Thus we find that
\be
r_{outer}<r_{0, outer}\,.
\ee
This is a violation of equation \eqref{sevenq}, since it implies
\be
1-{\mu G M(r_{outer})\over r^{d-2}_{outer}}= 1-{r_{0,outer}^{d-2}\over r_{outer}^{d-2}}<0\,.
\ee

\item $\Delta r \gtrsim \Delta r_{crit}$\,:

The argument here will be the same as the one for the case $T_\text{ECO}<T_\text{H}$. The redshift at the ECO surface is, using (\ref{kone})
   \be
 q(R_\text{ECO})\approx  {r_0^\h\over (d-2)^\h (R_\text{ECO}-\t r_0)^\h}\lesssim {r_0^\h\over  (\Delta r_{crit})^\h}\sim \left ( {r_0\over l_p}\right)^{d-1\over d+1}\, .
 \ee
 This violates the condition ECO2  (eq.\,(\ref{ksixq})) which requires that we have a higher redshift at $R_\text{ECO}$.
 
 Thus we find that for $T_\text{ECO}>T_\text{H}$, we cannot get an ECO satisfying the conditions ECO1-ECO3.

\end{enumerate}

\section{Using the Tolman-Oppenheimer-Volkoff equation}\label{sec6}

In the analysis of the previous section, we ignored the feedback of the radiation on the metric. In this section we will make a more self-consistent analysis of the field equations near the surface  $r=R_\text{ECO}$. 

We have taken our metric to be described by the spherically symmetric ansatz (\ref{five}) in the region $r>R_\text{ECO}$.  In the near surface region the local temperature is very high, due to the large redshift required by condition ECO2. Thus a typical quantum field with mass $m\ll m_p$ will appear massless to leading order. We will assume the stress tensor of the thermal gas near the ECO surface to have the form of a perfect fluid. With the directions of the orthonormal frame in  aligned along $t,r,$ and the angular directions. we have
\be
T^\mu{}_{\nu}{}^{\rm radiation}={\rm diag}\{ -\rho^{\rm r}(r), p^{\rm r}(r), p^{\rm r}(r), \dots, p^{\rm r}(r) \}\,,
\label{zone}
\ee
where $\rho ^{\rm r}(r)$ and $p^{\rm r}(r)$ denote the density and pressure of the radiation respectively. Due to the large  redshift factor in this region and the correspondingly high local temperature, 
we choose the equation of state 
\be
p^{\rm r}(r)={1\over d} \rho^{\rm r}(r)\,,
\label{ztwo}
\ee
 appropriate to a  massless field. 

In the vacuum black hole geometry, in the high redshift region near $r=r_0$, the vacuum energy near the horizon also has the form (\ref{zone}),(\ref{ztwo}). One can see this from the picture of the near horizon Boulware vacuum as the  Rindler vacuum; the vacuum stress-tensor of this Rindler vacuum is the negative of the stress tensor of a thermal gas at temperature $T={1\over 2\pi s}$, and such a thermal gas satisfies (\ref{zone}),(\ref{ztwo}). 

Recall from the discussion of section \ref{sec3} that vacuum energy near the ECO surface has the same order of magnitude as the energy density of thermal quanta. We will assume that the vacuum energy just outside $r=R_\text{ECO}$ also has a form like (\ref{zone}),(\ref{ztwo})
\be
T^\mu{}_{\nu}{}^{\rm vacuum}={\rm diag}\{ -\rho^{\rm v}(r), p^{\rm v}(r), p^{\rm v}(r), \dots, p^{\rm v}(r) \}\, ,
\label{zonew}
\ee
where $\rho ^{\rm v}(r)$ and $p^{\rm v}(r)$ denote the density and pressure of the vacuum respectively. Further we assume 
\be
p^{\rm v}={1\over d} \rho^{\rm v}\, .
\label{ztwow}
\ee

We cannot provide a rigorous  justification for  this assumption, but we will give arguments to make it plausible.  Since the metric outside the ECO is not the metric of a black hole, a priori we have to solve the analog of equation (\ref{mfour}) for the field modes in the ECO geometry, and then compute the vacuum stress energy from those modes. This is a difficult computation, and so we assume (\ref{zonew}) as a heuristic extrapolation motivated by the properties of the vacuum stress tensor in the black hole geometry. It is important to note that we are not specifying the {\it value} of $\rho^{\rm v}(r)$ for the vacuum stress tensor; we are just assuming that the vacuum energy stress tensor is diagonal, isotropic and traceless. In the next section we will work with the 1+1 dimensional system, where we will know the vacuum energy explicitly in the fully backreacted geometry of the ECO. For now we just check that (\ref{zonew}) is consistent to leading order with the requirements of the vacuum stress tensor.

Consider for example a  massless scalar field in 3+1 dimensions. The anomaly $\langle T^\mu{}_\mu \rangle$ has terms which are order curvature squared; e.g., we have a term  $R^{\mu\nu}R_{\mu\nu}$ with coefficient of order unity (see for example \cite{Birrell:1982ix}).  From Einstein's equations $G_{\mu\nu}=8\pi G T_{\mu\nu}$, we see that 
\be
R^{\mu\nu}R_{\mu\nu}\sim G^2 \rho^2\, .
\ee
We compare this contribution to $\langle T^\mu{}_\mu \rangle$ to the component $\rho$ of the stress tensor, finding
\be
{G^2 \rho^2\over \rho} \sim {G^2\rho}\sim {l_p^4 T^4}\sim {l_p^4\over s^4}\, ,
\ee
where we have noted that the temperature $T_{\text{ECO}}(r)$ of the ECO  at a proper distance $s$ from the horizon radius goes like $T\sim {1\over s}$. (Recall that we are always assuming that $T_{\text{ECO}}\sim T_\text{H}$.) So, as long as we look at distances $s\gg l_p$, this contribution to the  trace $\langle T^\mu{}_\mu \rangle$ is much smaller than the components $\rho^{\rm v}, p^{\rm v}$ of the stress tensor. 

Similar arguments give bounds on other contributions to the anomaly $\langle T^\mu{}_\mu \rangle$. For example
\be
\square R \sim {1\over s^2} GT^\mu{}_\mu \lesssim {G\rho\over s^2} \sim {l_p^2\over s^6}\, ,
\ee
Here we have replaced the derivatives in $\square$ with $1/s^2$, with $s$ being the scale over which all quantities vary in the near horizon region, and in the second step used the fact that the trace $T^\mu{}_\mu$ must be less than or order $\rho$. Again we find
\be
{\square R\over \rho} \lesssim {l_p^2\over s^2}\,,
\ee
which is small for $s\gg l_p$.

We therefore see that it is  consistent to assume a traceless vacuum stress tensor to leading order, as encoded in the condition (\ref{ztwow}). 
Assuming (\ref{zonew}), (\ref{ztwow}) for the vacuum stress energy, the total stress energy (radiation +vacuum) in the near-surface region also has the form
\be
T^\mu{}_{\nu}{}={\rm diag}\{ -\rho(r), p(r), p(r), \dots, p(r) \}\, ,
\label{eq.Ttotal}
\ee
with the pressure 
\be
p={1\over d} \rho\, .\label{eq.eostotal}
\ee
 Our goal now is to solve the Tolman-Oppenheimer-Volkoff (TOV)  equations with such a stress tensor in the region just outside $r=R_\text{ECO}$. 

\subsection{Approximating the TOV equation}

From the estimates in section \ref{sec5}, we note that we are interested in the situation where the stress tensor in the region $r>R_\text{ECO}$ falls off rapidly  with increasing $r-r_0$. Since $R_\text{ECO}$ is very close to the value $r_0$, we will be able to use the approximation
\be
r\approx  r_0\,.
\label{ttwo}
\ee
in many of the terms in the TOV equation. This approximation will allow us to simplify the equation. We will find that this approximate equation has no solution which yields an ECO. We will then
re-examine the approximations we made in simplifying the TOV equation, and find that these approximations become invalid only when we are outside the compactness scale $s_c$ defined in (\ref{zzel}).  Thus we will conclude that if an ECO satisfies  condition ECO1, then there must be vanishing (at leading order) of the stress tensor (\ref{eq.Ttotal}) at $r>R_\text{ECO}$.  As we saw in section\,\ref{sec3},  the vanishing of this stress tensor happens when  $T_\text{ECO}=T_\text{H}$ and the geometry in the region just outside $r>R_\text{ECO}$ becomes the geometry of the traditional black hole with a vacuum around the horizon.  

Before proceeding to obtain the approximate form of the TOV equation valid near $r\approx r_0$, we note what is known about the exact TOV equation, for the case where the stress tensor is given by  (\ref{eq.Ttotal}), (\ref{eq.eostotal}). The Einstein equations can be reduced to a single nonlinear second order equation called the Emden-Chandrasekhar equation \cite{chandrasekhar}. This equation has an analytically known  solution which is singular at the origin. For the $3+1$ dimensional case, the solution has the form.
\be
\rho={Q_1\over r^2}, ~~~ e^{2\alpha}=Q_2 r, ~~~e^{2\beta}=Q_3\,,
\label{jone}
\ee
where $Q_i$ are positive constants. The solution regular at the origin cannot be solved for analytically, but its asymptotic form has been developed as a power series. The singular solution is discarded when we consider the context of stellar structure, where the `truncated isothermal sphere' can be used to model the core of a star. In our problem, we have the opposite situation, where the region $r<R_\text{ECO}$ is a quantum gravitational region {\it not} described by isothermal semiclassical physics, while the region $r>R_\text{ECO}$ can take the form of an isothermal region, truncated at some radius $r_{max}$. Thus the singular  solution (\ref{jone})  is also an allowed solution in the region $R_\text{ECO}<r<r_{max}$. 

This singular solution shows the relevance of our condition ECO3, which says that there should not be too much matter far from the ECO surface.   Take the singular isothermal solution (\ref{jone}), truncated at some large radius $R$.  Choose a small radius $r=\epsilon$, and replace the singular region in its interior by some unspecified quantum dynamics.  Now if we let $R_\text{ECO}=\epsilon$, we see that, formally, such a solution will satisfy the conditions ECO1, ECO2. Condition ECO1 says that $r_\text{ECO}$ be not much larger than $r_0$, and here we have $R_\text{ECO}=\epsilon\ll r_0$. Condition ECO2 requires a high redshift at $r=R_\text{ECO}$, and from(\ref{jone}) we see that $q(\epsilon)=e^{-\beta(\epsilon)}$ is indeed large. But note that almost all the  mass of this object  is outside $r=R_\text{ECO}$. To exclude such a  case from our analysis we had imposed the condition ECO3.

 While we cannot solve the isothermal TOV equation exactly, we will find that  we have a high energy density in the region very close to $r=r_0$. In this situation we can get an approximation to the TOV equation that we can solve in closed form, and this solution will yield a more rigorous derivation of the heuristic estimates of section \ref{sec5}.

\subsubsection{Preliminary steps}

For the metric ansatz (\ref{five}), the conservation law $T^{r\mu}{}_{;\mu}=0$  gives 
\be
{\alpha '}=-{ p'\over p+\rho}\, ,
\ee
where a prime denotes ${d\over dr}$. Setting $p=\frac{\rho}{d}$ from  (\ref{eq.eostotal}) gives us
\be
{\alpha'}= -{1\over (d+1)} {\rho'\over \rho}\, .
\ee
The solution to this equation is
\be
e^{2\alpha}={C_2\over |\rho|^{\frac{2}{d+1}}}\, ,
\label{eq.gtt}
\ee
where $C_2>0$ is a constant. 

The Einstein equation $G_{tt}=8\pi G T_{tt}$ gives the $g_{rr}$ coefficient $e^{2\beta(r)}$ through (\ref{ktwo}),(\ref{ktwop}). From (\ref{ktwop}) we find 
\be
M'(r)=\Omega_{d-1} r^{d-1} \rho(r)\,.
\label{tfour}
\ee
With the approximation (\ref{ttwo}), this becomes
\be
M'(r)\approx \Omega_{d-1} r_0^{d-1} \rho(r)\,.
\label{tfourq}
\ee

\subsubsection{Approximating  the TOV equation}

The TOV equation in $d+1$ dimensions reads 
\beq
r^{d-1}p'(r) = -{(d-2)\mu\over 2}GM(r)\rho(r)\left(1+\frac{p(r)}{\rho(r)}\right)
\left ( 1+{16\pi  p(r)\,r^d\over \mu (d-2)(d-1)M(r)} \right ) 
\left(1-\frac{\mu G M(r)}{r^{d-2}}\right)^{-1}\, .
\label{tDthree}
\eeq
We simplify this equation using the approximation (\ref{ttwo}):

\begin{enumerate}[label=(\alph*)]

\item

Using our assumed equation of state $p={1\over d}\rho$ we get
\be
-r^{d-1} p'(r)= -{1\over d} r^{d-1} \rho'(r)\approx -{1\over d} r_0^{d-1} \rho'(r)\,.
\ee

\item

We have
\be
1+{p(r)\over \rho(r)}= {d+1\over d}\,.
\ee

\item 

We have
\be
1+{16\pi  p(r)\,r^d\over \mu (d-2)(d-1)M(r)}=1+{16\pi  \rho(r)\,r^d\over \mu d(d-2)(d-1)M(r)}\approx 1+{16\pi  \rho(r)\,r_0^d\over \mu d(d-2)(d-1)M(r)}\,.
\ee
We also note that if we were to distribute a mass $M$ uniformly over a radius $r_0$, we would get a density \be
\rho_{\text{uniform}}\sim  {M\over r_0^d}\,.
\ee
By contrast, in the example of section \ref{secoutline}, we had taken $s_\text{ECO}\sim l_p$, and found that the energy of radiation in a planck width shell was also order $\sim M$. So, in that example the energy density $\rho(r)$ near the ECO would be higher than $\rho_{\text{uniform}}$ by a factor $r_0/l_p\gg 1$. More generally, we assume that the energy density near the surface of the ECO is much higher than $\rho_{\text{uniform}}$, which implies that
\be
p\sim \rho \gg {M\over r_0^d}\,.
\label{ztw}
\ee
 Thus we make the approximation
 \be
 1+{16\pi  \rho(r)\,r_0^d\over \mu d(d-2)(d-1)M(r)}\approx {16\pi  \rho(r)\,r_0^d\over \mu d(d-2)(d-1)M(r)}\,.
 \ee
 
\item

We have
\be
\left(1-{\mu GM(r)\over r^{d-2}}\right)^{-1}\approx \left(1-{\mu GM(r)\over r_0^{d-2}}\right)^{-1}\,.
\ee

\end{enumerate}

\b

With these approximations, and using (\ref{tfourq}) to express $\rho$, the TOV equation (\ref{tDthree})  reduces to
\beq
 \Big(1-\frac{\mu GM(r)}{r_0^{d-2}}\Big)M''(r) = -\frac{\mu G}{2r_0^{d-2}}\frac{(d+1)}{ d }(M'(r))^2\,,\label{tDfive}
\eeq
In what follows we will call this  the approximate TOV equation.

\subsubsection{Solving the approximate TOV equation}

Let us write 
\be
u(r)=1-{\mu GM(r)\over r_0^{d-2}}\, .
\label{zseven}
\ee
Then (\ref{tDfive}) becomes
\beq
u(r) u''(r) = \frac{(d+1)}{2d}\left(u'(r)\right)^2 \,.
\eeq
The solution to this equation is
\beq
u'(r)= C_3\, u(r)^{\frac{d+1}{2d}}\,,
\eeq
and a final integration gives
\beq
u(r)=\left(\frac{(d-1)}{2d}C_3(r-r_1)\right)^{\frac{2d}{d-1}}\,.
\eeq
Thus we have
\beq
1-\frac{\mu GM(r)}{r_0^{d-2}}=\left(\frac{(d-1)}{2d}C_3(r-r_1)\right)^{\frac{2d}{d-1}}\,.
\label{tDseven}
\eeq

\subsubsection{Range of validity of the approximation}

We have obtained the approximate TOV equation (\ref{tDfive}) for the near-surface region of the ECO. This approximation will be good for a range
\be
R_\text{ECO}<r\lesssim r_{max}\,.
\label{eq.rmax}
\ee
We will now obtain an estimate for  $r_{max}$.

\begin{enumerate}
    \item  Recall that in eq. (\ref{ztw}) we had set
\be
{p(r)r_0^d\over M(r)}\gg 1\,.
\label{zthire}
\ee
This inequality is expected to hold close to the ECO surface, since $p(r)\sim\rho(r)\sim T_\text{ECO}^{d+1}(r)\sim T_\text{H}^{d+1}(r)$, and $T_\text{H}(r)$ is large near the ECO surface due to the large redshift $q(r)$. 

To get a rough estimate of scales, we use the expression (\ref{kone}) for $q(r)$ for the black hole geometry to get
\be
p(r)\sim T_\text{H}^{d+1} q(r)^{d+1} \sim {1\over r_0^{d+1}} \left ( { r_0\over r-r_0}\right)^{d+1\over 2}\sim {1\over r_0^{d+1\over 2} (r-r_0)^{d+1\over 2}}\,.
\label{zfive}
\ee
We also have
\be
M(r)\sim M\sim {r_0^{d-2}\over G}={r_0^{d-2}\over l_p^{d-1}}\,.
\ee
Thus,
\be
{p(r)r_0^d\over M(r)}\sim  {l_p^{d-1}\over \left(r-r_0\right)^{d+1\over 2}  r_0^{\frac{d-3}{2}}}\,.
\ee
We find that ${p(r)r_0^d\over M(r)}\sim 1$ at 
\be
(r-r_0)\sim \left ({l_p^{d-1}\over r_0^{d-3\over 2}}\right )^{2\over d+1}\,.
\label{zzsixt}
\ee
Using the relation (\ref{knine}) to write this value of $r-r_0$ in terms of the proper distance $s$ from $r_0$ in the black hole metric, we find
\be
s\sim M^{2\over (d-2)(d+1)} l_p\,.
\label{ccttwo}
 \ee
Thus, we find that for the condition (\ref{zthire}) to hold, the value of $r_{max}$ defined in (\ref{eq.rmax})  is given by the same distance scale $s_c$ (eq.\,(\ref{zzel})) that was used to define the compactness of the ECO  in condition ECO1.

\item

In (\ref{tDfive}) we needed to evaluate the derivative of ${ M(r)\over r}$,
but we approximated this by the derivative of ${ M(r)\over r_0}$; thus we were ignoring the variation of the value of $r$ in the near surface region. Writing
\be
{d\over dr} \left ({ M(r)\over r}\right ) ={1\over r}{dM(r)\over dr}-{M(r)\over r^2}\,,
\ee
we see that this approximation is valid as long as the ratio between the two terms on the RHS satisfies
\be
z(r)\equiv \left ( {M(r)\over r^2}\right) \left ( {1\over r}{dM(r)\over dr}\right )^{-1}\ll1\,.
\ee
Using (\ref{tfour}) and the same estimates for $\rho(r)$ that we used for $p(r)$ in (\ref{zfive}), we find
\be
z(r)\sim {(r-r_0)^{d+1\over 2} M\over r_0^{d-1\over 2}}\,.\ee
Thus $z(r)\sim 1$ at
\be
(r-r_0)\sim \left ( {r_0^{d-1\over 2}\over M}\right )^{2\over d+1}\sim \left ({l_p^{d-1}\over r_0^{d-3\over 2}}\right )^{2\over d+1}\, .
\label{cctone}
\ee
This is the same distance scale that appeared (\ref{zzsixt}), so we again find that $r_{max}$ is the radius given  by the compactness scale $s_c$.  

\end{enumerate}

Thus in the discussion below, we will set
\be
r_{max}-r_0 =\left ({l_p^{d-1}\over r_0^{d-3\over 2}}\right )^{2\over d+1}\,.
\ee
Since the radius $r\sim r_{max}$ is at the compactness scale $s_c$, we see by  condition ECO3 that the redshift factor at $r_{max}$ must be of the same order as the redshift factor predicted by the black hole geometry. Thus
\be
q(r_{max})\sim \left ( {r_0\over l_p}\right)^{d-1\over d+1}\sim \left( {M\over m_p}\right )^{{(d-1)\over (d-2)(d+1)}}\,.
\label{zzthree}
\ee

\subsection{Analyzing the approximate TOV solution}

First, consider the power ${2d\over d-1}$ appearing on the RHS of  (\ref{tDseven}). We are considering $d\ge 3$. For $d=3$, this power is an odd integer $3$, while for all higher $d$ it is a fraction. The LHS of (\ref{tDseven}) must be a positive real number by the requirement (\ref{eq.ECO4}), so we need
 \be
 C_3(r-r_1)>0\,,
 \label{zzsevent}
 \ee
 throughout the domain (\ref{eq.rmax}) where our approximate solution is valid. This gives us two possibilities, which we study in turn.

\subsubsection{The case $C_3>0, ~r_1<R_\text{ECO}$:}

 One possibility is that we have  $C_3>0,$ and $r_1<R_\text{ECO}$, which gives (\ref{zzsevent}) throughout the domain (\ref{eq.rmax}). 
 Using (\ref{tDseven}) we find
\beq
\rho(r) = -\frac{d}{8\pi G r_0} \left(\frac{(d-1)}{2d}|C_3|\right)^{\frac{2d}{d-1}} (r-r_1)^{d+1\over d-1}\,.
\label{tDsix}
\eeq
We see that $\rho(r)<0$. Thus this situation is similar to the case $T_\text{ECO}<T_\text{H}$ which had $\rho(r)<0$ in our heuristic analysis of section \ref{sec5}. From (\ref{eq.gtt}) we find
\be
e^{2\alpha(r)}={C_2\over |\rho|^{\frac{2}{d+1}}}=C_2\left (\frac{d}{8\pi G r_0} \left(\frac{(d-1)}{2d}|C_3|\right)^{\frac{2d}{d-1}}\right) ^{-\frac{2}{d+1}}{  (r-r_1)^{-\frac{2}{(d-1)}}}\,.
\label{teight}
\ee
Note that $e^{2\alpha(r)}$ determines the redshift factor
\be
q(r)=(-g_{tt}(r))^{-\h}= e^{-\alpha(r)} = C_2^{-\h}\left (\frac{d}{8\pi G r_0} \left(\frac{(d-1)}{2d}|C_3|\right)^{\frac{2d}{d-1}}\right) ^{\frac{1}{d+1}}{ (r-r_1)^{\frac{1}{d-1}}}\,.
\label{teightz}
\ee
We see  that since $r>r_1$, the redshift factor $q(r)$ {\it increases} as we move $r$ to larger values. Thus $q(r)$ will keep increasing monotonically until at least the location $r=r_{max}$ where our near-surface approximation (\ref{ttwo}) to the TOV equation fails.   Thus
\be
q(R_{ECO})<q(r_{max})\sim \left( {M\over m_p}\right )^{{(d-1)\over (d-2)(d+1)}}\,,
\ee
where in the second step we have recalled (\ref{zzthree}).  Thus we see that we do not satisfy the high redshift condition (\ref{ksixq}).

\subsubsection{The case $C_3<0, ~r_1>R_\text{ECO}$:}

Now consider the possibility that
\be
 C_3<0, ~~~~r_1>R_\text{ECO}\, .
 \ee
 We write $\rho(r)$  as
\beq
\rho(r) = \frac{d}{8\pi G r_0} \left(\frac{(d-1)}{2d}|C_3|\right)^{\frac{2d}{d-1}} (r_1-r)^{d+1\over d-1}\,.
\label{tDsixp}
\eeq
We see that $\rho(r)>0$. Thus this situation is similar to the case $T_\text{ECO}>T_\text{H}$ which had $\rho(r)>0$ in our heuristic analysis of section \ref{sec5}.

 Note that the LHS of (\ref{tDseven}) is required to not vanish anywhere, but the RHS vanishes at $r=r_1$. We avoid an inconsistency only if $r_1$ is sufficiently large that values $r\approx r_1$ lie outside the range where the near-surface approximation (\ref{ttwo}) of the TOV equation is valid. Recall that $r_{max}$ is the same order as the compactness scale $s_c$, and that outside this compactness scale  the standard black hole geometry is expected to be a reasonable approximation to the geometry. Noting that the solution   (\ref{tDseven}) is a power law in the near-surface region, we state the requirement that $r_1$ be significantly outside the range $r_0<r<r_{max}$ by requiring
  \be
r_1-r_{max} \gtrsim r_{max}- r_0\,.
\label{zel}
\ee
 Even though we say that $r_1$  is far outside the compactness scale $s_c$,  we will take $r_1-r_0\ll r_0$, and  will use this approximation to simplify some relations; taking a larger $r_1$ does not change the argument that follows.

Again, using (\ref{eq.gtt}) we have
\be
q(r)=(-g_{tt}(r))^\h = e^{-\alpha(r)} = C_2^{-\h}\left (\frac{d}{8\pi G r_0} \left(\frac{(d-1)}{2d}|C_3|\right)^{\frac{2d}{d-1}}\right) ^{\frac{1}{d+1}}{ (r_1-r)^{\frac{1}{d-1}}}\,.
\label{teightq}
\ee
In the standard black hole geometry, 
\be
q(r)\approx \left({(d-2)(r-r_0)\over r_0}\right)^{-\h}\,.
\label{xxel}
\ee
The expression (\ref{teightq}) holds for $R\lesssim r_{max}$ while by condition ECO3, the expression (\ref{xxel}) holds for $r\gtrsim r_{max}$. Thus at $r\sim r_{max}$ both expressions should have approximately the same value. Thus we must have
\be
C_2^{-\h}\left (\frac{d}{8\pi G r_0} \left(\frac{(d-1)}{2d}|C_3|\right)^{\frac{2d}{d-1}}\right) ^{\frac{1}{d+1}}{ (r_1-r_{max})^{\frac{1}{d-1}}} \approx \left({(d-2)(r_{max}-r_0)\over r_0}\right)^{-\h}\,.
\ee
Thus the  constants $C_2, C_3$ in the above relation satisfy
\be
C_2^{-\h}\left (\frac{d}{8\pi G r_0} \left(\frac{(d-1)}{2d}|C_3|\right)^{\frac{2d}{d-1}}\right) ^{\frac{1}{d+1}}\approx \left({r_0\over (d-2)(r_{max}-r_0)}\right)^\h{ 1\over  (r_1-r_{max})^{\frac{1}{d-1}}}\,.
\ee
Then the redshift at $r=R_\text{ECO}$ is given by (using eq.(\ref{teightq}))
\be
q(R_\text{ECO})\approx \left({r_0\over (d-2)(r_{max}-r_0)}\right)^\h
\left ( { r_1-R_\text{ECO}\over  r_1-r_{max}}\right )^{\frac{1}{d-1}}\,.
\label{zzeightt}
\ee
We now wish to use the intuition that  $r_1$ is `large' in the sense  (\ref{zel}). We have
 \be
 {r_1-R_\text{ECO}\over r_1-r_{max}}=1+{r_{max}-R_\text{ECO}\over r_1-r_{max}}<1+{r_{max}-r_0\over r_1-r_{max}}<1+{r_{max}-r_0\over r_{max}-r_{0}}=2 \,,
 \ee
 where in the last step we have used the  relation $(r_1-r_{max})\gtrsim (r_{max}-r_0)$. Then (\ref{zzeightt}) gives
 \be
q(R_\text{ECO})\sim \left(r_0\over (d-2)(r_{max}-r_0)\right)^\h \sim  \left(\frac{r_0}{l_p}\right)^\frac{d-1}{d+1}\,.
\ee
Since condition ECO2 requires a redshift much larger than this at the ECO surface (eq.\,(\ref{ksixq})), we conclude that we again do not find an acceptable solution to the approximate TOV equation (\ref{tDfive}).

\subsection{Checking the consistency of condition ECO3}\label{seccheck}

The goal of condition ECO3 was to exclude situations where there is a significant amount of matter outside the radius $R_{\text{ECO}}$; having such matter would conflict with the notion that our object is `extremely compact'. But we have noted that there is a certain amount of stress energy outside $R_{\text{ECO}}$ that is unavoidable, since this stress-energy results from the state of quantum fields at temperature $T_{\text{ECO}}(r)$  in the near-surface region. We should therefore check that the stress-energy of this thermal gas is low enough at $s\gtrsim s_c$ so that we can indeed require that the geometry in this region is close to the black hole geometry. In this section we will check that such is the case, using estimates from the analysis above. 

For the ECO geometry (\ref{five}), the metric coefficient $e^{2\beta(r)}$  is given by (eq.\,\ref{ktwo}))
\be
e^{-2\beta(r)}=1-{\mu G M(r)\over r^{d-2}}
\label{cctthree}
\,,\ee
with
\be
M(r)=M-E_{rad}(r)\,.
\ee
Here $E_{rad}(r)$ is the energy contributed by the radiation  in the region $r<r'\lesssim 2r_0$, where (as in the above sections) we have truncated our thermal gas at a scale $2r_0$. Following the lines of the computation leading to  (\ref{eq.Erad}) we have
\be
E_{rad}(r)\sim {r_0^{d-3\over 2}\over (r-r_0)^{d-1\over 2}}\,.
\ee
In the black hole metric we have
\be
e^{-2\beta_\text{H}(r)}=1-{\mu G M\over r^{d-2}}\,.
\ee
Then we have, (still assuming $r\lesssim 2r_0$)
\be
{e^{-2\beta(r)}\over e^{-2\beta_\text{H}(r)}}=1+{\mu G E_{rad}(r)\over r^{d-2}-\mu GM}=1+{\mu G E_{rad}(r)\over r^{d-2}-r_0^{d-2}}\approx 1+{\mu G E_{rad}(r)\over (d-2)r_0^{d-3}(r-r_0)}\,.
\ee
Using arguments similar to the ones leading to (\ref{eq.Erad}) to estimate $E_{rad}(r)$, we find
\be
{\mu G E_{rad}(r)\over (d-2)r_0^{d-3}(r-r_0)} \sim { l_p^{d-1} \over r_0^{d-3\over 2}(r-r_0)^{d+1\over 2}}\,.
\ee
For the ECO geometry to approximate the black hole geometry we need that the above quantity be $\lesssim 1$. This yields the requirement
which is
\be
(r-r_0)\gtrsim \left ({l_p^{d-1}\over r_0^{d-3\over 2}}\right )^{2\over d+1}\,.
\ee
From (\ref{zzsixt}),(\ref{ccttwo}) we see that this is just the condition $s\gtrsim s_c$. Thus the coefficient of $g_{rr}$ in the ECO geometry satisfies the constraint ECO3.

The coefficient $\alpha(r)$ in the ECO geometry (\ref{five}) is given through the equation
\be
{(d-1)\alpha'(r)e^{-2\beta(r)}\over r}=- {(d-1)(d-2)(e^{-2\beta(r)}-1)\over 2r^2}+{8\pi G p(r)}\,.
\ee
Using (\ref{ktwo}), this is
\be
{(d-1)\alpha'(r)e^{-2\beta(r)}\over r}= {(d-1)(d-2)\mu G M(r)\over 2r^d}+{8\pi G p(r)}\,.
\label{cctfive}
\ee
For the black hole, we have (i) $e^{-2\beta_\text{H}(r)}=1-{\mu G M\over r_0^{d-2}}$, (ii) $M(r)=M$,  and (iii) $p=0$. Then we get
\be
\alpha'_\text{H}(r)={(d-2)\mu G M\over 2r^{d-1}(1-{\mu GM\over r^{d-2}})}\,,
\ee
which gives
\be
e^{2\alpha_\text{H}(r)}=1-{\mu G M\over r^{d-2}}\,.
\ee

From the discussion above we have seen that $e^{-2\beta(r)}$ can be approximated by its black hole value for $s\gtrsim s_c$. On the RHS, again using estimates for $E_{rad}$ similar to the one in (\ref{eq.Erad}), we find that for $s\gtrsim s_c$, 
\be
1-{M(r)\over M}={E_{rad}(r)\over M} \sim {l_p^{d-1}\over r_0^{d-1\over 2} (r-r_0)^{d-1\over 2}} \sim {l_p^{d-1}\over s^{d-1}} \lesssim {l_p^{d-1}\over s_c^{d-1}} \sim \left ( {l_p\over r_0}\right)^{2(d-1)\over d+1}\ll 1\,.
\ee
Thus in the first term on the RHS of (\ref{cctfive}) we can write $M(r)\approx M$. To see if $\alpha(r)\approx \alpha_\text{H}(r)$, we just need to check that the last term on the RHS of (\ref{cctfive}) can be ignored compared to the first term on the RHS; i.e., we need
\be
{p(r)r^d\over M}\lesssim 1\,.
\ee 
But using the analysis leading to (\ref{zzsixt}), we find  that this inequality holds for $s\gtrsim s_c$.

To summarize, the thermal energy of quantum fields in the region $s\gtrsim s_c$ is small enough that the geometry in this region can be approximated by the black hole geometry. In particular, in this region the redshift in the ECO geometry is of the same order as  its value in the black hole geometry. Thus it is consistent to impose condition ECO3   in our definition of an ECO to disallow additional sources of stress-energy outside $R_{\text{ECO}}$.

\subsection{Summary}

Let us summarize the discussion of this section. In section\,\ref{sec5} we had performed a heuristic analysis of the total stress-energy --- vacuum plus radiation --- in the near-surface region  of an ECO. We found that if $s_{\text{ECO}}<s_c$, then the backreaction of this stress-energy leads to a singularity in the metric (\ref{five}), in the semiclassical region $r>R_{\text{ECO}}$. The analysis required us to separately consider the cases  $T_\text{ECO}<T_\text{H}$ and $T_\text{ECO}>T_\text{H}$, and led to the conclusion that the only allowed situation is the one with $T_\text{ECO}=T_\text{H}$, where the stress-energy in this near-surface region vanishes. This analysis was not completely self-consistent however, as we used the metric of the ``core'' of the ECO to find the redshift in the near-surface region, and did not take into account the fact that  this redshift would also be affected by the stress-energy outside the core. Since the redshift at a point determines the energy density of radiation at that point, our estimates were  heuristic to the extent that they did not use a fully self-consistent redshift. 

In the present section we remedied this difficulty  by considering the TOV equation which describes the  metric coupled self-consistently with the stress-tensor. We gave plausibility arguments for adopting the perfect fluid stress-tensor (\ref{eq.Ttotal}), (\ref{eq.eostotal}). Eq.\,(\ref{eq.gtt}) links the energy density at a point with the redshift at that point, now in a self-consistent fashion. Again we found that there is no regular solution satisfying the properties required of an ECO, if $s_{\text{ECO}}<s_c$. The analysis splits into two cases, which correspond to negative and positive values of $\rho$; thus these cases parallel the $T_\text{ECO}<T_\text{H}$ and $T_\text{ECO}>T_\text{H}$ cases of section\,\ref{sec5}. We conclude that the only situation with no singularity in the semiclassical region $r>R_\text{ECO}$ is the one with $\rho=p=0$, which we identify with the  situation $T_\text{ECO}=T_\text{H}$ following (\ref{vvone}).

\section{An ECO in 1+1 dimensional dilaton gravity}\label{sec7}

In our study of the near-surface dynamics of an ECO, an important role has been played by the contribution to the stress-tensor from the quantum vacuum near the ECO surface. In the analysis of section \ref{sec6}, we  allowed the vacuum energy density $\rho^{\text{v}}(r)$ to be arbitrary, but used some heuristic arguments to assume the form  (\ref{eq.Ttotal}),(\ref{eq.eostotal}) for the overall form of  the stress tensor. In this section we will consider the case of 1+1 dimensional dilaton gravity, coupled to massless scalars. In this setting we can compute the vacuum energy explicitly using the conformal anomaly, and thus get a set of self-consistent equations describing the region $r>R_\text{ECO}$. More precisely, there will be vacuum energy  as well as radiation energy outside the ECO, and the vacuum energy will be consistent with the  metric which results from these two energies.

\subsection{The model}

The CGHS model \cite{PhysRevD.45.R1005} describes the quantum dynamics of 1+1 dimensional dilaton gravity coupled to a set of massless scalars $f_i$. The action is
\begin{gather}
    I_\text{CGHS} = \frac{1}{2\pi}\int d^2x\sqrt{-g}  [e^{-2\phi}(R+4(\nabla\phi)^2+4\lambda^2)-\sum_{i=1}^n\frac{1}{2}(\nabla f_i)^2]\,.
\end{gather}
Integrating out the scalars $f_i$ gives a a Polyakov term in the effective action of the form $R\square^{-1}R$. This nonlocal term can be cast in a local form using an auxiliary field $\psi$ with action
 \begin{equation}
     I_1 = -\frac{\kappa}{2\pi}\int d^2 x \sqrt{-g} (\frac{1}{2}(\nabla\psi)^2+\psi R)\,.
     \label{ztfive}
 \end{equation}
Here $\kappa$ is related to the number of scalar fields by $\kappa=(n-24)/24$. The shift $n\r n-24$ is due to ghosts, and we will assume that $n>24$ so that $\kappa>0$.  
The dynamics of these coupled fields can be simplified if we modify the action by the addition of a term $ I_2 = -\frac{\kappa}{2\pi}\int d^2 x \sqrt{-g} \phi R$, which yields the RST model \cite{PhysRevD.46.3444}. We then arrive at the action
\begin{equation}
   I= - \frac{1}{2\pi}\int d^2x\sqrt{-g}\Big[ e^{-2\phi}(R+4(\nabla\phi)^2+4\lambda^2)+\kappa\Big(\frac{(\nabla\psi)^2}{2}+\psi R+\phi R\Big)\Big]\,.
   \label{jfive}
\end{equation}
Variation of the metric $g^{\mu\nu}$ gives
 \begin{align}\label{eq.totEMT}
\frac{e^{-2\phi}}{\pi}\Big(
-2g_{\mu\nu}(\square\phi-(\nabla\phi)^2+&\lambda^2)+2\nabla_{\mu}\nabla_\nu\phi\Big)-\frac{\kappa}{2\pi}\Big(\nabla_\mu\psi\nabla_\nu\psi-2\nabla_\mu\nabla_\nu\psi\nonumber
\\
&-g_{\mu\nu}(-2R +\frac{1}{2}(\nabla\psi)^2)\Big)-\frac{\kappa}{\pi}(g_{\mu\nu}\square\phi-\nabla_\mu\nabla_\nu\phi)=0\,.
 \end{align}
Variation of the dilaton $\phi$ gives
 \begin{equation}\label{eq.dileq}
e^{-2\phi}(R-4(\nabla\phi)^2+4\square\phi+4\lambda^2)= -\kappa\frac{R}{2}\,,
 \end{equation}
 and variation of the auxiliary field $\psi$ gives
 \begin{equation}\label{eq.psieq}
     \square\psi = R\,.
 \end{equation}
Taking the trace of the metric equation, and using the other two field equations, we find
\be
(R+2\square\phi)(\frac{\kappa}{2}-e^{-2\phi})=0\,.
\ee
We are interested in the solution where $\phi$ is not a constant, which implies
 \begin{equation}
     R = -2\square\phi\,.
 \end{equation}
Combined this with \eqref{eq.psieq} gives
 \begin{equation}
 \label{eq.weq}
     \psi = -2\phi + w \quad\implies\quad  \square w =0\,.
 \end{equation}
 
\subsection{Solving the equations}
We wish to solve the above equations using a coordinate choice which is similar to the Schwarzschild type coordinates (\ref{five}) that we used in higher dimensions. 
We list below key steps from the discussion of  \cite{PhysRevD.105.025015}, where the above equations were nicely analyzed.
\begin{enumerate}[label=(\roman*)]
    \item We take the static ansatz
 \begin{equation}
   ds^2=-\tilde{g}(x)dt^2+\tilde{g}^{-1}(x)dx^2=-g(\phi)dt^2+g^{-1}(\phi)h^2(\phi) d\phi^2\, ,  
   \label{zfift}
 \end{equation}
 where in the second step we have used $\phi$ as a spatial coordinate in place of $x$, and $\t g(x)=g(\phi(x))$. At spatial infinity we will have flat spacetime, with curvature $R=0$. From eq.(\ref{eq.dileq}) we find that $\phi=\pm\lambda x$ at infinity; we choose the solution $\phi=-\lambda x$. At $x\r\infty$ we will then get 
 \be
 \phi\r -\infty, ~~~~g(\phi)\r 1, ~~~~h(\phi)\r -{1\over \lambda}\,.
 \label{zninet}
 \ee 
For the above static metric, the equation $\square w=0$ \eqref{eq.weq} can be solved: 
\begin{equation}
\partial_x w(x)=\frac{C}{\tilde{g}(x)} \,  \quad \implies \quad   \partial_\phi w(\phi)\equiv w'=C\frac{h(\phi)}{g(\phi)}\,,
\label{eq.Cintro}
\end{equation}
where $C$ is an integration constant, and primes denote derivatives with respect to $\phi$. 

\item In the metric (\ref{zfift}), the $\phi\phi$ component of the metric equation \eqref{eq.totEMT} gives, after some algebra
\beq
\left(1-\frac{\kappa}{2}e^{2\phi}\right)\frac{h'(\phi)}{h(\phi)}=-\kappa e^{2\phi}\left(\left(1-\frac{Ch(\phi)}{2g(\phi)}\right)^2+\frac{C}{2}\frac{g'(\phi)h(\phi)}{(g(\phi))^2}\right)\,.
\label{ztwenty}
\eeq
The dilaton equation \eqref{eq.dileq} gives
\beq\label{dileqn}
-\left(1-\frac{\kappa}{2}e^{2\phi}\right)\left(\frac{h'(\phi)}{h(\phi)}-\frac{g'(\phi)}{g(\phi)}\right)-2+2\lambda^2\frac{(h(\phi))^2}{g(\phi)}=0\, .
\eeq
For the static metric (\ref{zfift}), $R=\p_x^2 \t g(x)$. Converting derivatives in $\p_x$ to derivatives $\p_\phi=h\p_x$, the equation  $R=-2\square\phi$
 can be integrated to give
\beq
g'(\phi)=2g(\phi)-d_\lambda \, h(\phi)\, ,
\label{zsixt}
\eeq
where $d_\lambda$ is an integration constant. Eq. (\ref{zsixt})  can be solved by introducing a new function $Z(\phi)$,
\begin{align}
g(\phi)=-\frac{d_\lambda}{2\lambda} \, e^{2\phi} Z(\phi), ~~ ~~~
h(\phi)=e^{2\phi}\frac{Z'(\phi)}{2\lambda}.\;\label{zeighttq}
\end{align}
The limit (\ref{zninet}) gives 
\be
d_\lambda=-2\lambda.
\ee
so that we have
\begin{align}
g(\phi)= e^{2\phi} Z(\phi), ~~ ~~~
h(\phi)=e^{2\phi}\frac{Z'(\phi)}{2\lambda}.\;\label{zeightt}
\end{align}

\item In the dilaton equation (\ref{dileqn}) we note that ${g'\over g}-{h'\over h}={h\over g}\left ( {g\over h}\right )'$.  Then (\ref{dileqn}) becomes
\be
\left(\left(e^{-2\phi}-\frac{\kappa}{2}\right)\left(\frac{g}{h}\right)\right)'=-2\lambda^2 h e^{-2\phi} = -\lambda Z'\,,
\ee
which can be integrated to yield
\be
g(\phi)=\frac{2\lambda h(\phi)e^{2\phi}}{\kappa e^{2\phi}-2}( Z(\phi)+A_C)\, ,
\label{zttwo}
\ee
where $A_C$ is an integration constant.  Substituting in this relation the expressions (\ref{zeightt}) for $g(\phi)$ and $h(\phi)$, we get an equation for $Z(\phi)$ which is easily integrated to give
\beq
Z(\phi)+A_C\ln Z(\phi)=\kappa\phi + e^{-2\phi}+a_1\, ,
\label{ztone}
\eeq
where $a_1$ is another integration constant. 

\item We must now relate the constant $A_C$ to the integration constant $C$ that we had introduced earlier in \eqref{eq.Cintro}. In the metric equation (\ref{ztwenty}) we can replace ${h'\over h}$ from  (\ref{dileqn}) and then replace any derivatives $g'$ using (\ref{zsixt}). This gives
\be
  2\lambda^2\frac{h^2}{g}+\frac{2\lambda h}{g}-\frac{\lambda h}{g}\kappa e^{2\phi} =  -\frac{C}{2}\left(\frac{C}{2} +2\lambda \right)\frac{h^2}{g^2}\kappa e^{2\phi}\,.
\ee
Using the forms (\ref{zeightt}) for $g$ and $h$, this gives
\be
 e^{2\phi}\frac{(Z')^2}{2Z}+\frac{ Z'}{Z}-\frac{Z'}{Z}\frac{\kappa}{2} e^{2\phi} =  -\frac{C}{2}\left(\frac{C}{2} +2\lambda \right)\frac{(Z')^2}{4\lambda^2 Z^2}\kappa e^{2\phi}\,.
 \label{ztthree}
 \ee
Also, using the forms (\ref{zeightt}) for $g$ and $h$ in (\ref{zttwo}) gives
\be
Z' = \frac{Z e^{-2\phi}}{Z+A_C}(\kappa e^{2\phi}-2)\,.
\ee
Using this in (\ref{ztthree})  gives
\begin{align}
-\frac{e^{-2 \phi} \left(\kappa e^{2 \phi}-2\right)^2 \left(-8 A_C \lambda ^2+C^2 \kappa+4 C \kappa \lambda \right)}{16 \lambda ^2 (A_C+Z)^2}&=0\,,
\end{align}
from which we read off
\begin{align}
A_C&=\frac{\kappa}{8\lambda^2}C(C+4\lambda)\, \nonumber \,.
\end{align}

\item Summarizing the above steps, we find that the complete solution of our coupled equations is given in terms of one integration constant $C$, in the form 
\beq
g(\phi)=e^{2\phi}Z(\phi) \, , \ \ \ h(\phi)=\frac{1}{2\lambda}e^{2\phi}Z'(\phi)\, ,
\label{zthone}
\eeq
where $Z(\phi)$ is found by solving the equation
\be
Z(\phi)+A_C\ln\left(\frac{Z(\phi)}{|A_C|}\right)=e^{-2\phi}+\kappa\phi -a_2 \, ,   ~~~~~A_C=\frac{\kappa}{8\lambda^2}C(C+4\lambda) \,,
\label{zthtwo}
\ee
where we have written $a_2=-a_1+A_C\log |A_C|$.

\end{enumerate}

\subsection{Finding the Boulware and Hartle-Hawking states}

The integration constant $C$ determines the different quantum states of the matter fields on our spacetime through their contribution to the stress-tensor at infinity. Recall that we have integrated out the matter fields to obtain a nonlocal Polyakov which has then been rewritten in terms of the auxiliary field $\psi$, giving the action (\ref{ztfive}). Varying $g^{\mu\nu}$ in this action gives the stress tensor
\be
T^{(1)}_{\mu\nu} =-\frac{\kappa}{2\pi}\Big(\nabla_\mu\psi\nabla_\nu\psi-2\nabla_\mu\nabla_\nu\psi-g_{\mu\nu}(-2R +\frac{1}{2}(\nabla\psi)^2)\Big)\,.
\label{zzfour}
\ee
Using the metric (\ref{zfift}) to compute covariant derivatives, and using the relations $\psi = -2\phi +w$ and $w' = \frac{Ch}{g}$,  we get
\be
T^{(1)}_{t}{}^{t} = -\frac{\kappa}{2\pi} \left(\frac{6g}{h^2}+\frac{4\lambda}{h}+\frac{C(C+4\lambda)}{2g}-\frac{4gh'}{h^3}\right)\,.
\label{ztseven}
\ee
At $x\r\infty$, we have the limit (\ref{zninet}), which yields
\begin{align}
 \rho(x=\infty)=-   T^{(1)}_{t}{}^{t}(x=\infty)= \frac{\kappa}{4\pi} (C+2\lambda)^2\,.
 \label{jthree}
\end{align}
The Boulware vacuum corresponds to the case of no energy density at infinity, so we need $  T^{(1)t}_{t}(x=\infty)=0$. This gives
\be
C=-2\lambda\,.
\ee
In the Hartle-Hawking vacuum we need smoothness at the horizon. This implies a a finite value of the energy density  at the horizon where $g\r0$.  In (\ref{ztseven}) we see that the third term on the RHS diverges as $g\r0$, unless
\be
C=0 ~~~{\rm or} ~~~ C=-4\lambda\,.
\label{zthir}
\ee
Thus these two values of $C$ describe the Hartle-Hawking vacuum. With the values of $C$ in (\ref{zthir}), we have $A_C=0$. The horizon corresponds to the point where $g_{tt}=-g(\phi)=0$, which from (\ref{zthone}) implies $Z=0$. From (\ref{zthtwo}), we see that the horizon is at $\phi=\phi_0$ where 
\be
e^{-2\phi_0}+\kappa\phi_0 -a_2=0\,,
\label{jfour}
\ee
We define 
\be
M={\lambda a_2\over \pi}\,,
\label{zztwo}
\ee
 as the mass of the hole.

\subsection{Analysis of the solution}

With the above solution of 1+1 dimensional gravity coupled to scalars, we can now address the temperature of an ECO. Let $T_\text{H}$ be the temperature of the black hole and thus the temperature at infinity of the Hartle-Hawking vacuum. Putting the values of $C$ from (\ref{zthir}) into (\ref{jthree}) we find that the energy density at infinity of the Hartle-Hawking vacuum is $\rho_{HH}(\infty)={\kappa \lambda^2\over \pi}$. For $T_\text{ECO}<T_\text{H}$, the energy density at infinity should be less than  $\rho_{HH}(\infty)$, which implies $-4\lambda<C<0$. For this range of $C$, we have $A_C<0$. Conversely, for $T_\text{ECO}>T_\text{H}$, we need $C<-4\lambda$ or $C>0$; in these cases we have $A_C>0$. 

Note that the radius of the black hole can be considered to be $r_0\equiv e^{-\phi_0}$ and since we are interested in a large black hole, we will take
\be
\phi_0\ll 0\,.
\ee
Thus $|\phi_0|\ll e^{-\phi_0}$; we can use this to simplify our qualitative understanding of the solution of (\ref{zthtwo}). 

Let us now consider the cases $T_\text{ECO}<T_\text{H}$ and $T_\text{ECO}>T_\text{H}$ in turn.

\subsubsection{$T_\text{ECO}<T_\text{H}$}\label{seclower}

In this case $A_C<0$, so we write $A_C=-|A_C|$. For ${(T_\text{H}-T_\text{ECO})\over T_\text{H}}\sim 1$, we have $|A_C|\sim \kappa$. We will assume $\kappa\sim 1$ in what follows, and thus take $|A_C|\sim 1$. 

At spatial infinity, we have $\phi\r -\infty$, and from (\ref{zthtwo}) we have $Z\sim e^{-2\phi}$. Let us move inwards from infinity, which implies that we move towards larger values of $\phi$, and look for the first point where we have a difficulty with the solution of (\ref{zthtwo}). Differentiating this equation we find
\be
Z'={(\kappa-2e^{-2\phi})\over (1-{|A_C|\over Z})}\,.
\label{jsix}
\ee
Thus the solution fails when we reach $Z=|A_C|$. From (\ref{zthtwo}) we see that this value of $Z$ corresponds to a point $\phi_*>\phi_0$. Using (\ref{jfour}) we find
\be
\Delta\phi \equiv (\phi_*-\phi_0)\approx -{|A_C|\over 2} e^{2\phi_0}\,.
\label{zzfift}
\ee
Since $\phi_0\ll 0$ for a large black hole, we see that $|\Delta \phi|$ is very small.  At $\phi=\phi_*$ we find that 
\be
g(\phi_*)=Z(\phi_*) e^{2\phi_*}\approx {|A_C| \over 2} e^{2\phi_0}\,.
\ee
Thus the redshift parameter at $\phi_*$ is
\be
q(\phi_*)\equiv (-g_{tt}(\phi_*))^{-\h}=(g(\phi_*))^{-\h}\approx \left ({2\over |A_C|}\right )^\h e^{-\phi_0}\,.
\ee
Again noting that  $\phi_0\ll 0$, we have $q(\phi_*)\gg 1$.

Let us now try to place this solution in the context of an ECO. Since the solution fails at $\phi= \phi_*$, we should place a surface just outside this point, at a location we call $\phi_\text{ECO}$ with $\phi_\text{ECO}<\phi_*$. We then imagine that some other quantum gravitational dynamics takes over in the region $\phi>\phi_\text{ECO}$. Of course we have already solved the full quantum gravity dynamics arising from the action (\ref{jfive}), with the solution given in (\ref{zthtwo}). Thus to get some novel dynamics at $\phi>\phi_{ECO}$, we must imagine that there are other quantum fields in the complete theory. These fields  can be, for example, fields with large mass that do not contribute significantly to the action at $\phi \ll 0$, but which can still modify the dynamics to something that is regular at $\phi>\phi_\text{ECO}$. Such a picture brings the problem to the same footing as our analysis of the situation in higher dimensions $d$.

We can now see qualitative similarities between what we have found above and the situation for higher $d$. Since we must take $\phi_\text{ECO}<\phi_*$, we find that the redshift at the ECO surface $\phi_\text{ECO}$ is bounded as
\be
q(\phi_\text{ECO})< \left ({2\over |A_C|}\right )^\h e^{-\phi_0}\,.
\label{zzone}
\ee
Thus we cannot have an ECO for $T_\text{ECO}<T_\text{H}$ if we demand that  the redshift at the ECO surface  is higher than the value in the RHS of  (\ref{zzone}). Note that this maximum redshift is finite but large, since $\phi_0\ll 0$.

We cannot make a more precise comparison to the situation for higher $d$ since we do not have any simple way to extrapolate expressions like (\ref{seven}) to the case $d=1$. This difficulty is already present at the classical level. For $d>2$, the radius of the hole grows with its mass as $r_0\sim M^{1\over d-2}$. Extrapolating this expression  to $d=1$ would suggest $r_0\sim M^{-1}$, which would indicate a radius that goes down as the mass increases. But from (\ref{zztwo}) we expect that  even for the case $d=1$,  a larger $M$ implies a larger radius $r_0\equiv e^{-\phi_0}$. We will find further differences between the cases of $d=1$ and higher $d$ below.

\subsubsection{$T_\text{ECO}>T_\text{H}$}\label{sechigher}

In this case $A_C>0$, so we write $A_C=|A_C|$ for clarity. Again, for ${(T_\text{ECO}-T_{H})\over T_\text{H}}\sim 1$, we have $|A_C|\sim \kappa$. As in the above subsection, we will  assume $\kappa\sim 1$ in what follows, and thus take $|A_C|\sim 1$. 

This time the analogue of (\ref{jsix}) is
\be
Z'={\kappa-2e^{-2\phi}\over 1+{|A_C|\over Z}}\,.
\label{jsixq}
\ee
We have $Z>0$ throughout our analysis, since $g$ would become singular at $Z=0$. The denominator is thus regular for all $Z$. But the numerator vanishes at 
\be
\phi_s=-\h \ln {\kappa\over 2}\,.
\ee
We see that $\phi_s$ is a number of order unity, and thus describes the `strong coupling' region of dilaton gravity. This location can be considered to be the analogue of   $r_s\sim l_p$ in higher dimensions. We are not interested in looking at the analogue of a central singularity at $r=0$, but at the behavior near the horizon location $\phi=\phi_0$. While we do not have a singularity in the solution at $\phi\approx \phi_0$, there is an interesting change of behavior as we pass this location,  which we now note.

For $\phi<\phi_0$, the RHS of (\ref{zthtwo}) is positive. At $\phi=\phi_0$, the RHS vanishes. For 
\be
\phi_s\ll \phi\lesssim \phi_0\,,
\label{jseven}
\ee
 the RHS becomes $\approx -a\sim - e^{-2\phi_0}$ (we have used $\phi_0\ll 0$ to drop terms like  $\phi_0$ compared to exponentials $e^{-2\phi_0}$). This large negative value is reproduced on the LHS by a very small (positive) value of $Z$. The first term $Z$ on the LHS becomes ignorable, and we are left with
\be
|A_C|\ln {Z\over |A_C|}\approx -e^{-2\phi_0}\,,
\ee
 which gives (using $|A_C|\sim 1$)
 \be
 Z\sim e^{-e^{-2\phi_0}}\,,
 \label{zzsix}
 \ee
 which is a very tiny number. From this we find that in the region (\ref{jseven})
 \be
 g(\phi) \sim e^{-e^{-2\phi_0}}e^{2\phi_0}\,.
 \label{zznine}
 \ee
The second factor on the RHS is a small correction to the first factor, and not relevant by itself in the approximation in which we have written (\ref{zzsix}), but we keep this factor in $g$ since we will be comparing $g$ to $h$ below, and the ratio will involve  this subleading factor.   We also have
 \be
 Z'={(\kappa-2e^{-2\phi})\over (1+{|A_C|\over Z})}\approx -{2\over |A_C|} e^{-2\phi_0} Z\,,
\label{jsixqq}
\ee 
and
\be
h(\phi)=\frac{1}{2\lambda}e^{2\phi}Z'(\phi) \approx \frac{1}{2\lambda}e^{2\phi_0}Z'(\phi)\approx -\frac{1}{2\lambda}{2\over |A_C|}  Z \sim -{1\over \lambda}e^{-e^{-2\phi_0}}\,.
\label{jsixqqq}
\ee 
Thus in the  metric (\ref{zfift}), in the region (\ref{jseven}),  the coefficient of $-dt^2$ and $d\phi^2$ are both very small due to the factor $ Exp[-e^{-2\phi_0}]$.  Note however that
\be
{g\over  \lambda h}\sim  e^{2\phi_0}\ll 1\,.
\label{zzeight}
\ee

Suppose one were to put a surface at some point $\phi_\text{ECO}$ in the region (\ref{jseven}) and replace its interior by some new physics. Then this object would satisfy the spirit of condition ECO1 since $\phi_\text{ECO}>\phi_0$ (i.e. the ECO surface is {\it inside} the horizon radius). The redshift parameter at this surface is
\be
q(\phi_\text{ECO})\equiv (-g_{tt}(\phi_\text{ECO}))^{-\h}\sim e^{\h e^{-2\phi_0}}\,,
\ee
which cannot be made arbitrarily large, but must still be considered extremely large since $\phi_0\ll 0$. Thus in a rough sense one could say that this looks like an ECO, with $T_\text{ECO}>T_\text{H}$. We can see how this situation escaped our earlier arguments against such an ECO in higher dimensions. On the RHS of eq.\,(\ref{tDseven}), for example,  we have the power ${2d\over d-1}$ which is a finite positive number for $d>1$, but diverges for $d=1$. A similar singularity $\sim {1\over d-1}$ in a power is encountered in other relations  like (\ref{tDsix}), (\ref{teightq}). Thus our general analysis does not work for the case $d=1$, and we may have an ECO like object with $T_\text{ECO}>T_\text{H}$.

\subsection{The form of the stress tensor}

Recall that in section\,\ref{sec6} we had given arguments as to why the vacuum stress tensor could be taken to be traceless to leading order in the high redshift region near the surface of an ECO. This had led to a traceless form of the total stress energy (eq.(\ref{eq.Ttotal})), (\ref{eq.eostotal}). In the present case of $d=1$, we have the exact stress energy from our solution, so we can check if such a traceless condition is indeed maintained. 

In the case of $T_{ECO}<T_\text{H}$ studied in section\,\ref{seclower}, the solution becomes singular outside the horizon radius, so we cannot approach closer than to the horizon than the distance $|\Delta\phi|$ given in (\ref{zzfift}). But in the case $T_{ECO}>T_\text{H}$ studied in section\,\ref{sechigher}, the solution continues past the horizon location $\phi=\phi_0$. Thus we can investigate if the condition $p\approx \rho$ holds in the region near the horizon in the exact analysis of the $d=1$ case.

From the stress tensor (\ref{zzfour}) we find
\be
T^{(1)}_{x}{}^x=p= \frac{\kappa}{2\pi}\left(\frac{28 g}{h^2}+\frac{24\lambda}{h}+\frac{C(C+4\lambda)}{2g}-\frac{16gh'}{h^3}\right)\,,
\label{zzfive}
\ee
and
\be
-T^{(1)}_{t}{}^t=\rho =\frac{\kappa}{2\pi} \left(\frac{6g}{h^2}+\frac{4\lambda}{h}+\frac{C(C+4\lambda)}{2g}-\frac{4gh'}{h^3}\right)\,.
\label{zzsixc}
\ee
We are interested in the relation between $\rho$ and $p$ in the horizon region $\phi\approx \phi_0$, where the coefficients of the metric change very sharply. 
In this region we find that both $g$ and $h$ become very small, but as noted in (\ref{zzeight}), $g$ is smaller than $\lambda h$ by a factor $\sim e^{2\phi_0}$. We also need to estimate
\bea
h'&=&{1\over 2\lambda} e^{2\phi} Z''(\phi)~+~ {1\over \lambda}{(\kappa e^{2\phi}-2)\over 1+{|A_C|\over Z}}\,,\nn
&=&{\kappa\over \lambda} {e^{2\phi}\over 1+{|A_C|\over Z}}+{1\over 2\lambda}{ (\kappa e^{2\phi}-2)\over (1+{|A_C|\over Z})^2} {|A_C|\over Z^2} Z'\,,\nn
&\approx& {\kappa\over \lambda |A_C|} e^{2\phi_0}Z +{2\over \lambda |A_C|^2} e^{-2\phi_0} Z\,,\nn
&\approx & {2\over \lambda |A_C|^2} e^{-2\phi_0} Z\,.
\eea
Using (\ref{zzsix}) and (\ref{zznine}), we find that 
\be
{\lambda h'\over g}\sim e^{-4\phi_0}\,.
\label{zzseven}
\ee
Using (\ref{zzeight}),(\ref{zzseven}) in (\ref{zzfive}) and (\ref{zzsixc}), we find that in each case  one term  dominates over the other terms giving
\be
p\approx  \frac{\kappa}{2\pi}\frac{C(C+4\lambda)}{2g}\,,
\label{zzfiveq}
\ee
and
\be
\rho \approx \frac{\kappa}{2\pi}\frac{C(C+4\lambda)}{2g}\,.
\label{zzsixq}
\ee
Thus we find that $p\approx \rho$ near the horizon radius $\phi_0$.

\section{Discussion}\label{sec8}

In this paper we have argued that Extremely Compact Objects (ECOs) must have the same thermodynamic properties --- temperature $T$, entropy $S$ and radiation rates $\Gamma[\{l\},\omega]$ --- as a semiclassical black hole of the same mass. 

This result is relevant for the following reason. The semiclassical hole has an appealing thermodynamics, which seems to be related to the existence of a horizon. But  a  horizon leads to a loss of quantum unitarity in the process of black hole evaporation. In string theory we find that black hole microstates are fuzzballs with no horizon. This resolves the information puzzle, but leaves us with a different question: if fuzzballs have no horizon, then is there any reason to expect that they reproduce the thermodynamic properties of the semiclassical hole?

Entropic arguments suggest that fuzzballs have a surface that is at a distance $s\sim l_p$ outside the horizon radius. ECOs have also been postulated in many other theories of gravity (for a discussion, see for example \cite{cardoso}). We have found that if the ECO surface is at a distance $s_\text{ECO}\ll (M/m_p)^{2\over (d-2)(d+1)}$ from the horizon radius $r_0$, then the temperature $T_\text{ECO}$ must agree with the Hawking temperature $T_\text{H}$,  to an accuracy which improves as $s$ is made smaller. The central aspect of the argument involved studying the Tolman-Oppenheimer-Volkoff equation in the near surface approximation. This near-surface region is filled with a  gas of radiation, whose temperature and  energy density are very high due to the large redshift at the ECO surface. A negative contribution to the energy density comes from the vacuum `Casimir' energy, which we argued must be the same to leading order as the negative energy density of the local Rindler vacuum. We find that if the sum of radiation and vacuum energies does not vanish, then the back-reaction of the near energy density on the geometry does not allow for a consistent time-independent solution. The two sources of energy cancel for $T_\text{ECO}=T_\text{H}$, giving the agreement of temperatures mentioned above. 

Once we have an agreement of temperatures between the ECO and the black hole, the agreement of entropies  follows from standard thermodynamics. The agreement of radiation rates is due to  the large redshift at the ECO surface. This redshift separates the thermal bath near the ECO surface from the region where the quanta must penetrate an effective potential to emerge at infinity. This separation  leading to an agreement of graybody factors with the semiclassical hole once we are given an agreement of temperatures.

Thus these results provide a satisfying closure to the fuzzball paradigm of black holes. The information paradox is resolved because no microstate has a horizon. But the elegant thermodynamics emerging from the semiclassical hole is still preserved, due to the agreements of temperature, entropy, radiation rates. 

We have not studied the dynamical processes which would bring an ECO to a temperature $T_\text{ECO}=T_\text{H}$ if we initially start with an object with $T\ne T_\text{H}$. But qualitatively, we envisage the following. If $T>T_\text{H}$, then the quantum structure at $r<R_\text{ECO}$ will grow in size by absorbing the high temperature radiation near the object's surface, until the temperature just outside this surface drops to a value which equals the Hawking temperature, and a time-independent ECO can exist. Conversely, if $T<T_\text{H}$, then the quantum structure near the surface  will shrink,  giving up its energy to heat the radiation near the surface; equilibrium will be reached when the temperature of this radiation equals $T_\text{H}$, allowing an ECO to exist.   

Our analysis suggests several avenues for future work. We defined $\eta_T=T_\text{ECO}/T_\text{H}$, and argued that if $\eta_T\ne 1$, then there would be no consistent solution to the near-surface Einstein equations. But $\Delta T \equiv T_\text{ECO}-T_\text{H}$ could still be be nonzero and parametrically smaller than $T_\text{H}$. It would be interesting to work out how the allowed range $\Delta T$ goes to zero as $s_\text{ECO}$ goes towards smaller values. 

It is possible to get a  completely self-consistent solution of the quantum field equations in the 1+1 dimensional case. We have noted qualitative similarities between this case and the situation in higher dimensions, though there were some differences as well due to a divergence $\sim {1\over d-1}$ in the power law behavior of quantities in the Tolman-Oppenheimer-Volkoff equation.

It would also be interesting to extend the present analysis for black holes with charge and rotation, particularly as we go towards charge and rotation values that are extremal.

 \section*{Acknowledgments}

This work is supported in part by DOE grant DE-SC0011726. We would like to thank for discussions Iosif Bena, Robert Brandenberger, Marcel Hughes, Pierre Heidmann, Brandon Manley,  Ted Jacobson, Juan Maldacena, Robert Mann, Geoff Penington and Nicholas Warner.

\bibliographystyle{utphys} 
\bibliography{ECOthermodynamics}

\end{document}